\documentclass[11pt]{article}
\usepackage{graphicx}

\makeatletter
\def\ps@top{\let\@mkboth\@gobbletwo
     \def\@oddhead{\rm\hfil\thepage\hfil}\def\@oddfoot{}
     \def\@evenhead{}\let\@evenfoot\@oddfoot}
\def\@bibsetup{\itemindent=-\leftmargin}
\def\@citesep{; }
\def\@cite#1#2{({#1\if@tempswa , #2\fi})}
\def\@biblabel#1{\hfill}
\def\thebibliography#1{\section*{References\markboth
 {REFERENCES}{REFERENCES}}\list
 {[\arabic{enumi}]}{\settowidth\labelwidth{[#1]}\leftmargin\labelwidth
 \advance\leftmargin\labelsep
 \usecounter{enumi}\@bibsetup}
 \def\newblock{\hskip .11em plus .33em minus -.07em}
 \sloppy
 \sfcode`\.=1000\relax}
\renewcommand{\section}{\@startsection {section}{1}{\z@}{-3.5ex plus -1ex minus 
    -.2ex}{2.3ex plus .2ex}{\centering\large\bf}}
\renewcommand{\subsection}{\@startsection{subsection}{2}{\z@}{-3.25ex plus
    -1ex minus -.2ex}{1.5ex plus .2ex}{\centering\bf}}
\makeatother
\newcommand{\kms}{\mbox{km\,\ensuremath{\rm{s}^{-1}}}}

\begin{document}
\begin{center}{\bf\Large The Composition of Comets} \\ [0.5in]
Anita L. Cochran$^1$, Anny-Chantal Levasseur-Regourd$^2$,Martin Cordiner$^3$, \\
Edith Hadamcik$^4$, J{\'e}r{\'e}mie Lasue$^5$, Adeline Gicquel$^{3,9}$, \\
David G. Schleicher$^6$, Steven B. Charnley$^3$, Michael J. Mumma$^3$, \\
Lucas Paganini$^3$, Dominique Bockel{\'e}e-Morvan$^7$, Nicolas Biver$^7$,
Yi-Jehng Kuan$^8$ \\ [1in]

1. The University of Texas, McDonald Observatory, Austin, TX USA \\
2. UPMC, LATMOS, Paris, France \\
3. NASA Goddard Space Flight Center, Greenbelt, Maryland, USA \\
4. UPMC, LATMOS, Guyancourt, France \\
5. UPS, IRAP, Toulouse, France \\
6. Lowell Observatory, Flagstaff, AZ, USA \\
7. LESIA, Observatoire de Paris, Meudon, France \\
8. National Taiwan Normal University, Taiwan, ROC \\
9. MPS, G{\"o}ttingen, Germany

\end{center}
\newpage
\begin{center}{\bf Abstract} \end{center}
This paper is the result of the International Cometary Workshop, held in 
Toulouse, France in April 2014, where the participants came together to
assess our knowledge of comets prior to the ESA Rosetta Mission. In this
paper, we look at the composition of the gas and dust from the comae
of comets.  With the gas, we cover the various taxonomic studies that have
broken comets into groups and compare what is seen at all wavelengths.
We also discuss what has been learned from mass spectrometers during flybys.
A few caveats for our interpretation are discussed.  With dust, much of
our information comes from flybys.  They include {\it in situ} analyses
as well as samples returned to Earth for laboratory measurements.
Remote sensing IR observations and polarimetry are also discussed.
For both gas and dust, we discuss what instruments the Rosetta spacecraft
and Philae lander will bring to bear to improve our understanding of
comet 67P/Churyumov-Gerasimenko as ``ground-truth" for our previous
comprehensive studies.
Finally, we summarize some of the inital Rosetta Mission findings.

\section{Introduction}
Comets are leftovers from when our Solar System formed.
These icy bodies were formed out of the solar nebula in the outer regions
of our Solar System. Many of these bodies were swept up and incorporated
into the giant planets. The remnants of planetary formation were either
ejected from the nascent Solar System, were gravitationally perturbed to
the Oort cloud or remained in reservoirs past the orbit of Neptune.
The comets we see today have undergone little change from their primordial
states. Comets, therefore, represent important objects to study in order to
determine constraints suitable for models of the early solar nebula.

Whipple (1950) \nocite{wh50} first described comets as ``dirty snowballs"
that are composed of a mixture of ices and dust. 
The mass is approximately equally divided into ices and dust.
As the comets approach the Sun,
they are heated and the ices sublime. Since comets are small (generally a
few to 10s of km), the resultant gas is not bound to the nucleus and
expands outwards into the vacuum of space. When it leaves the nucleus, the
gas carries with it some of the solid particles (refractory grains,
refractory organics, icy grains, agregates of ice and dust).
The resultant material forms the cometary comae (and sometimes
tails) that are so prominent when a comet is in the inner Solar System.
We do not observe the nucleus composition directly, except when observing
from a spacecraft flying past or during a rendezvous mission.
What we generally observe is the gas in the coma and the light
reflected off the refractories and we {\it infer} composition from those.

The gases that come directly from the nucleus first flow through a
region near the nucleus where the gas densities are sufficiently high that
collisions, and thus chemical reactions, can take place.
This so-called ``collisional" zone is
$<1000$\,km for all but the most productive comets. With typical outflow
velocities near 1\,{\sc au} of 850 m/sec, this means that the outflowing
gas clears this region in the first 20 minutes of outflow. Once out of the
collisional zone, photochemical processes change the
composition of the gases.

The comets were formed over a wide range of heliocentric distances and
conditions. Petit {\it et al.} (2015, this volume) discuss the dynamics of the
formation.
In this review, we will discuss what we have learned about the chemical
composition of the gas and dust using studies at all wavelengths. Our
conclusions will be tempered by some caveats based on some current
observations.  The subject of isotope ratios, however, will be covered
in the review by Bockel{\'e}e-Morvan {\it et al.} (2015, this volume).
Finally, we will discuss what observations the Rosetta Mission can
make to improve our understanding, We include some early results of this
mission.

\section{Gas Composition}
The spectra of comets  are mostly composed of emissions from gas-phase molecules
superposed onto a continuum that results from sunlight reflecting off dust.
Many of the emissions arise from resonance fluorescence of the gas,
but prompt emissions also play an important role.
Optical spectra of comets have been obtained since the 1860s,
while UV, radio and IR spectra have been obtained since the
1970s. 

Optical spectra of comets contain few parent
species. The spectra are mostly bands of fragment molecules (often called daughters
though they can be grand-daughters or chemical products). Thus, the parent
species must be inferred using lifetimes and chemical reaction networks.
The IR and radio spectra (including mm and sub-mm) contain a mixture of parent
and daughter species. Many of the molecules only possess transitions
in the IR because they do not have permanent dipole moments.
In the UV, we see parents and daughters, as well as a number of important
atomic features, such as H, C and O.

Early on, it was noted that the spectra of different comets seemed
similar (rarely was an unusual feature noted in the spectra). The 
strengths of the emission lines relative to the continuum did seem to vary
and the relative strengths of bands changed with heliocentric distance.
This led to the question of whether all comets shared the same
composition or whether there were distinctly different classes of comets.
This question is important for our understanding of the homogeneity of
the solar nebula at the epoch(s) and the region(s) where comets formed.
Thus were born some large comparative studies of comets at all wavelengths
in order to understand the chemical homogeneity or diversity of comets.
Of course, even if comets formed identically, their spectra might have
been altered by activity or irradiation over their lifetime.
Thus, we need to look for
clues of evolution in the spectra.

\subsection{Optical Observations}
Since optical observations of comets have existed for far longer than spectra at
other wavelengths, there are many more comets observed in the optical than
at any other wavelengths. 
The optical observations consist of high-resolution
and low-resolution spectra, as well as photometry obtained with narrow-band
filters that isolate the molecular bands.
Optical observations are also the most sensitive to faint comets, meaning
that the optical observations can be gathered for more comets and
at larger heliocentric distances than the other wavelength observations.
Additionally, the typical photometric aperture or long slit used
is larger than those used in the IR, allowing observations
of a larger portion of the coma.
The larger the aperture, the less sensitive are the observations to
outflow velocity or short term temporal variability (which instead can
be measured at other wavelengths).
Feldman {\it et al.} (2004) discuss how to turn optical observations
of fragment species into knowledge of the composition of comets.
\nocite{fecoco04}

The spectral
observations have the advantage that they can easily isolate the molecular
features from the continuum, but suffer from small apertures.  Some observers
use long-slit spectrographs that allow for measurements of the coma as a 
function of cometocentric distance. This is an improvement over small apertures
and allows direct measurement of scale lengths for decay, but 
the long slit only allows for
sampling in specific directions within the coma. Spectroscopic observations
often have significant spectral grasp, so that multiple species can be
observed simultaneously, minimizing worries of temporal variability.
Large surveys with low-resolution optical spectra have been carried out
since the 1970s (Newburn and Spinrad 1984, 1989; Cochran 1987; Fink and Hicks
1996; Fink 2009; Langland-Shula and Smith 2011; Cochran {\it et al.}
2012) with well over 150 comets observed spectroscopically.
\nocite{nesp84,co87corr,fihi96,fi09,lasm11,coetal2012,nesp89}

High-resolution studies have tended to focus more on individual cometary
properties and are excellent for detailed work.  Examples of the types
of studies performed are $^{12}$C/$^{13}$C 
(Danks {\it et al.} 1974; Kleine {\it et al.} 1995),
$^{14}$N/$^{15}$N
(Arpigny {\it et al.}. 2003; Manfroid {\it et al.} 2005, 2009)
and O ($^1$S/$^1$D) (Festou and Feldman, 1981; Cochran and Cochran 2001;
Cochran 2008; Capria {\it et al.} 2005, 2008; McKay {\it et al.} 2012, 2013,
Decock {\it et al.}, 2013, 2015).
\nocite{dalaar74,kletal95a,aretal2003,maetal05,manfroidetal09}
\nocite{coco01,co08,fefe81,caetal2005,caetal2008,mckay12,mckay2013}
\nocite{decocketal2013,decocketal2015}

Photometry with filters
allows for much larger apertures and is therefore much more sensitive than
spectroscopy (Schleicher and Farnham 2004).
Continuum removal is done with observations through
filters that isolate continuum regions. Large photometric studies of
comets have been carried out (A'Hearn and Millis 1980; A'Hearn {\it et al.}
1995; Schleicher and Bair 2014), with more than 160 comets observed.
Photometry suffers from the problem of not being
able to measure weak features (e.g. CH or NH$_{2}$) because there
is a practical limit to how accurately the underlying continuum can be 
determined and removed.
\nocite{scfa04,ahmi80,ahetal95,scba14}

With photometry,
the different filters are observed sequentially, thus temporal
variations could change our understanding of the relative abundance of species.
However, the cycle time to observe all filters is typically much shorter
than the expected timescale for temporal variability.
With long-slit spectra, the spectra may be obtained at multiple
position angles to sample any inhomogeneities of the coma. At that point,
temporal variations must be considered. Typically, the time between
position angles is much longer than between filters with photometry.

Without prior knowledge of any asymmetry, single-aperture
photometry and small-aperture and long-slit spectroscopy all must
assume that the distribution of the gas in the coma is symmetric. This is
rarely the case. Examples of asymmetry can be seen in 
Cochran {\it et al.} (2012, see Figure 5).
While not able to probe the morphology of the coma, these observations
are, however, measuring the bulk composition of the nucleus. It is this
composition that is the necessary measurement to aid our understanding
of the conditions in the solar nebula, while morphology tells us more
about the physical condition of the nucleus today.
However, morphology also may tell us about inhomogeneities
of the nucleus, lending clues to the size and variation in composition
of the presumed cometesimals from which the nucleus was assembled.

To some extent, this assumption of symmetry can be mitigated with an integral
field unit (IFU) spectrograph, where spectra can be obtained from many
positions in the coma at the same time and with spectral resolution
sufficient to isolate bands. 
IFU spectroscopy is not as readily available as other instrument types and
most IFU spectrographs have very small fields-of-view.
Alternatively, the coma can be imaged with narrow-band filters to 
measure the morphology of the coma.

Early studies of cometary diversity were carried out by A'Hearn and Millis
(1980), Newburn and Spinrad (1984) and Cochran (1987).
\nocite{ahmi80,nesp84,co87corr}
These papers included measurements of daughters such as CN, C$_{2}$, C$_{3}$, 
and O\,I.  All the authors noted that the ratios of species appeared
very similar for most comets but there were occasional comets that 
seemed depleted in C$_{2}$ or C$_{3}$.  However, the sample sizes in these
first papers were small and it was not apparent what cometary properties were 
common to the unusual comets.

Since 1995, several large studies have been published with the total
number of comets studied by all groups around 250 (A'Hearn {\it et al.}. 1995;
Fink 2009; Langland-Shula and Smith 2011; Cochran {\it et al.} 2012;
Schleicher and Bair 2014).
Many of these comets were observed by more than one group.
Additional species were added in these surveys, including OH, NH, NH$_{2}$,
and CH.
\nocite{ahetal95,fi09,lasm11,coetal2012,scba14}
From these studies, it has become apparent that the majority of comets 
really do have very similar relative compositions. However, 25--30\% of
the comets look different than the rest.  As noted in the earlier studies,
many of the comets that are different are depleted in C$_{2}$ and C$_{3}$.

Unfortunately, intercomparing surveys is difficult, since the various
authors have used different $g$-factors and scale lengths in
their reductions, as well as different 
cut-offs for the definition of depletion. 
However, the surveys are mostly consistent in
comet-by-comet comparisons when they do overlap on the same
comet, agreeing which comets show differences.
The inconsistent reductions, however, are why ever larger surveys by a
single group can make the most progress.

\begin{figure}
\centering
\includegraphics[scale=0.5]{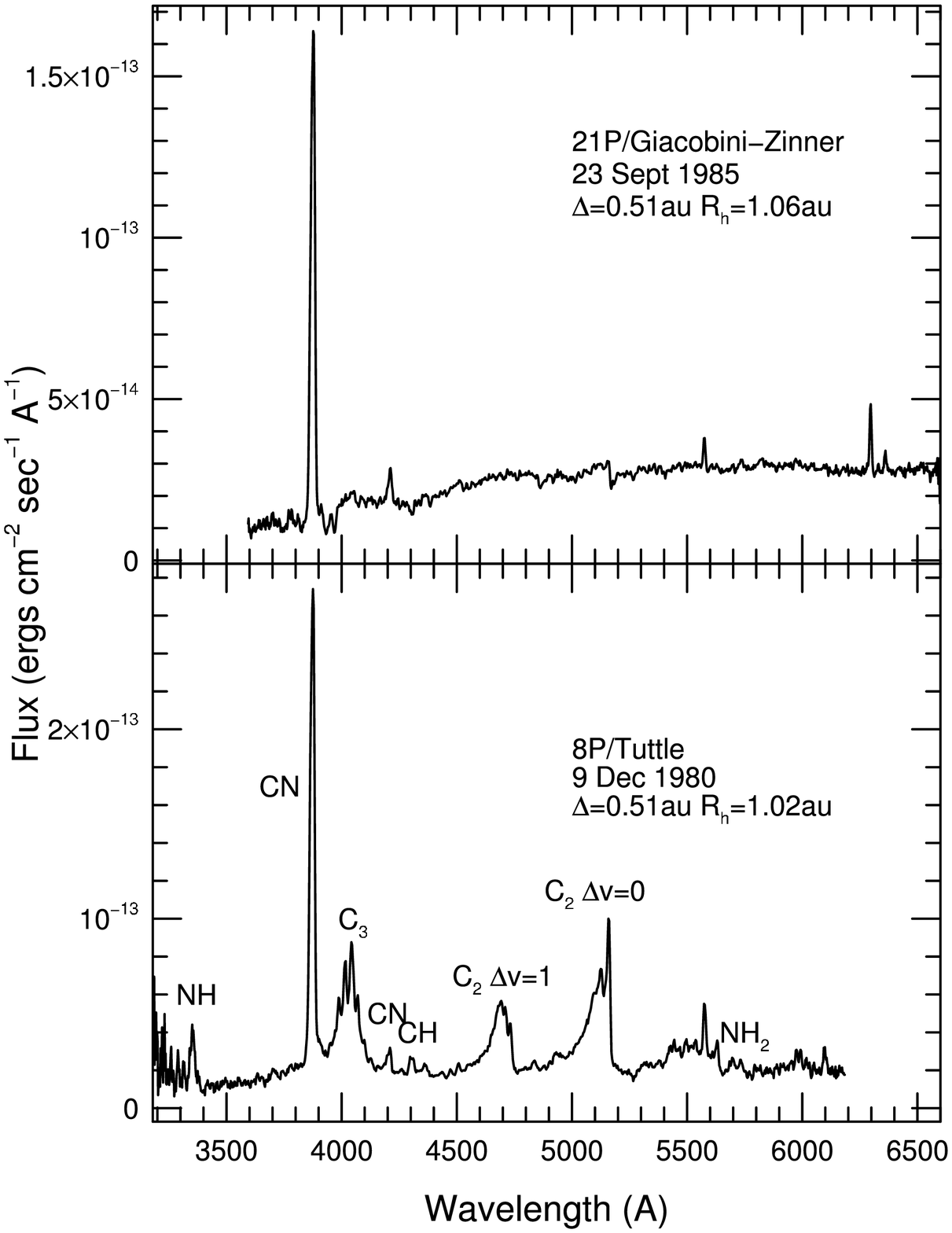}
\caption{Spectra of 21P/Giacobini-Zinner and 8P/Tuttle obtained at
McDonald Observatory are shown, scaled to the CN band at 3880\AA.
The comets were observed at comparable heliocentric and geocentric distances.  
Inspection of the figure shows that the well-defined C$_{2}$ and C$_{3}$ bands
seen in Tuttle are nearly absent in Giacobini-Zinner's spectrum.
(Spectra courtesy of A. Cochran)}\label{gz}
\end{figure}

Figure~\ref{gz} demonstrates that the depletions are not a subtle effect
by comparing
observations of comets 21P/Giacobini-Zinner (GZ) and 8P/Tuttle when both
were near 1\,{\sc au} from the Sun and 0.5\,{\sc au} from Earth.
The Tuttle spectrum demonstrates the full, normal, range of observed
molecules.  GZ is the prototypical depleted comet and it
is clear that there is much less C$_{3}$/CN and C$_{2}$/CN in GZ
than in Tuttle.

After the earliest studies, it seemed that the depletion occurred primarily in 
Jupiter Family comets. However, the studies of A'Hearn {\it et al.} (1995),
Fink (2009), Langland-Shula and Smith (2011) and Cochran {\it et al.} (2012)
had sufficiently large samples that it became apparent that the relationship
between dynamical state and depletion was not so simple. These studies
showed that depleted comets can be from any dynamical type: 
Jupiter Family, Halley Type, Long-period (dynamically new and not).
However, it is more common for depleted comets to be Jupiter
Family comets; Cochran {\it et al.} (2012) found that two thirds of the
depleted comets were Jupiter Family while one third were long period.
In addition, a higher percentage of the Jupiter Family comets (37\%)
were depleted than of the long period comets (18.5\%).
Thus, the depletion is either pointing to different 
evolutionary states or it is pointing to mixing of the formational
reservoirs.

Cochran {\it et al.} (2012) pointed out that the picture is a little
more subtle than just ``carbon-chain depleted" or ``typical".
The definitions for carbon-chain depletion typically only account
for C$_{2}$ relative to OH or CN and it is these C$_{2}$-depleted 
comets that represent 25--30\% of all comets. However, Cochran {\it et al.}.
found that only $\sim10$\% of comets are depleted in both C$_{2}$ 
\underline{and} C$_{3}$.  They also found some number of comets
with different patterns for some other molecules, but the statistics
did not warrant breaking the comets into additional groups.

Schleicher and Bair (2014) have analyzed photometry of 167 comets
observed and reduced in a consistent manner.  
Of that sample, 101 were of sufficient quality that they were able
to be included in the analysis of the statistics of the sample.
They find the
typical carbon-chain depletions found by other authors.  However,
their statistics are better than previous studies and they now
claim that there are 7 classes of comets: 1)~typical (still $\sim70$\%
of comets); 2)~comets depleted in both C$_{2}$ and C$_{3}$ with
the depletion as strong as GZ or stronger; 3)~comets
depleted in both C$_{2}$ and C$_{3}$ but weaker than the GZ depletion;
4)~comets depleted in C$_{2}$ but not C$_{3}$ (in agreement with
Cochran {\it et al.} 2012); 5)~comets depleted in C$_{3}$ but not
C$_{2}$; 6)~comets depleted in NH but not depleted in any carbon species;
and 7)~comets depleted in CN relative to C$_{2}$ and C$_{3}$ but still
depleted in C$_{2}$ and C$_{3}$ compared to OH
(a class of 1 -- 96P/Machholz~1; Schleicher 2008).
\nocite{sc08}

With the most recent findings, especially that certain depletions
are not confined to a single dynamical type, the question of
formation vs. evolution is important.  One of the best constraints
on this question is comet 73P/Schwassmann-Wachmann 3 (SW3).
In 1995, comet SW3 underwent a splitting event into three pieces; 
subsequently, those pieces split again.  SW3 is a strongly C$_{2}$ and C$_{3}$
depleted comet
and thus has a distinctive ``fingerprint". If the depletion were
just an evolutionary effect, from multiple perihelion passages,
we would expect it to be confined
mostly to the surface and the interior would appear typical.
However, observations of the distinct pieces during the 2006
apparition showed that all the pieces had identical depletions both
in the parents observed in the IR and the daughters observed in
the optical and IR (Kobayashi {\it et al.} 2007; Jehin {\it et al.} 2008;
Schleicher and Bair 2011).
In addition, there was no change from measurements of SW3 obtained
before the splitting in 1995 (Schleicher and Bair 2011).
This has been interpreted as strong
evidence that C$_{2}$ and C$_{3}$ depletions are primarily from
the formation of the
comets, not from their subsequent evolution.
\nocite{koetalsw3,scba11,jeetalsw3}

\subsection{UV Observations}
The UV part of the spectrum of comets is not as well studied
as the optical because the Earth's atmosphere precludes observations
from ground-based telescopes (the OH (0,0) band at 3080\AA\ is technically
in the UV but can be observed from the ground; however the atmospheric
extinction at this wavelength is typically around $\sim1.5$~mags/airmass).
Thus, UV observations require a spacecraft-borne telescope or 
a rocket experiment.
The electronic bands of some diatomic molecules are visible in the FUV; atomic
daughter products that result from solar UV dissociation are also seen.

In the UV, transitions of CO, H, H$_2$, O, C, OH, CS and S$_2$ have
been regularly detected.  Most of these species are products of
a parent molecule, though the processes to produce them are dependent
on the transition. 
Whether S$_2$ is a parent or a daughter is still unknown.; regardless
S$_2$ is very short-lived.
CO is seen to have two different band series,
the fourth positive bands and the Cameron bands.  
The fourth positive bands are the result of solar pumped fluorescence of CO. 
The Cameron bands result from either electron impact on CO or
dissociative excitation of CO$_2$, the latter being more important.
Thus, these two bands can yield information on two possible parents.

The UV era of cometary observations started in 1970 with the
Orbiting Astronomical Observatory (OAO-2) observations of comet 
Bennett (Code {\it et al.} 1972).  \nocite{coholi72}
In these observations OH and Ly $\alpha$, both daughters of
H$_{2}$O, were detected.

Significant progress in our understanding of the UV spectra of comets
came with the International Ultraviolet Explorer (IUE) satellite.
Ultimately, IUE was used to observe more than 50 comets between
1978 and 1996. Festou (1998) summarized the findings from these
observations.  He concluded that ``with the exception of the S$_2$
molecule, new and long-period comets have a chemical composition
that does not differ by more than a factor of two from that of
periodic comets" (but see discussion below).  \nocite{feIUE98}

Since 1990, the UV has been covered with the Hubble Space Telescope (HST) and
a variety of instruments.  Also operational during this period,
the Far Ultraviolet Spectroscopic Explorer (FUSE) mission operated from
1999 -- 2007 and the Galaxy Evolution Explorer (GALEX) operated
from 2003.  Of these three missions, HST has been the workhorse,
having been used to observe many comets.

FUSE was used to observe the FUV from 900 -- 1200\AA\ in four
comets. In that bandpass,
emissions due to H\,I, O\,I, CO (including three
new Hopfield-Birge band systems), H$_2$ and N\,I were detected.
A He\,I resonance transition at 584\AA\ was seen in second order
and was taken as an indicator of solar wind charge transfer in the coma.
Ar and N$_2$ were searched for but not detected 
(Feldman 2005; Feldman {\it et al.} 2009).
In addition, FUSE detected many lines that were initially unidentified,
but were later determined to be H$_2$ (Liu {\it et al.} 2007).
\nocite{fe2005,feetal2009,liuetal2007}

The GALEX satellite uses time-tagged images and grism mode to obtain both
FUV and NUV observations.  The FUV observations have concentrated on C\,I.
The NUV is used to observe OH, CS and CO$^+$ (Morgenthaler {\it et al.} 2009).
\nocite{moetal09,moetal11}
Comets fill the field-of-view of the GALEX detector so it is difficult to
estimate the background.  Morgenthaler {\it et al.} (2011)
describe the process of extracting the C\,I lifetime from observations
of comet C/2004~Q2 (Machholz).  They found that either the CO lifetime is
shorter than previously thought or another shorter-lived species must
be contributing to the production of C\,I in the inner coma.

Using the FIMS/SPEAR instrument on the Korean STSAT-1, Lim {\it et al.} (2014)
observed comet C/2001~Q4 (NEAT) in the FUV.  They observed the CO fourth
positive bands as well as C\,I and S\,I atomic features.  They concluded
that the production rate of C is $\sim60\%$ of CO, agreeing with 
results obtained for other comets, e.g.
Morgenthaler {\it et al.} (2011) for comet C/2004~Q2 (Machholz).  
For comet NEAT, Lim {\it et al.} found that S\,I production was about 1\%
of the water production rate.  It was assumed the parent was H$_2$S.
\nocite{limetal2014}

Meier and A'Hearn (1997) had already published a study of S\,I in 19
comets observed with IUE and HST.  They also observed CS in these spectra,
presumably a daughter of CS$_2$.  
Typical values for CS$_2$ derived are not sufficient to supply
all of the S.
Meier and A'Hearn pointed out that all the transitions they see are optically
thick. This means that simple fluorescence models are not sufficient and
they developed a radiative transfer/collisional model to interpret
their spectra.  For the comets they studied, Meier and A'Hearn derived
sulfur abundances of 0.1\% to 1\% relative to water.
\nocite{meah97}

HST has observed 18 comets with a variety of instruments. The comets
are a mixture of Jupiter Family and long-period comets.
Comet 103P/Hartley~2 (the target of the EPOXI mission) was observed
with HST during three different apparitions (1991, 1998, 2010), the only comet
observed over more than one apparition.  With most of these
observations, the CO fourth positive bands and the OH bands are
observed simultaneously, thus enabling a measure of the CO:H$_{2}$O 
for many comets. 
Lupu {\it et al.} (2007) presented models of the CO fourth positive
bands in four comets observed with HST.
These observations showed a large range
of Q$_{CO}$/Q$_{H_2O}$, from $<1\%$ for C/2000~WM1 (LINEAR)
to $>20\%$ for comet C/1996~B2 (Hyakutake).

Since then, Weaver {\it et al.} (2011) have reported that comet 103P/Hartley~2
had a CO/H$_{2}$O production rate ratio of $0.15-0.45\%$, amongst
the most carbon monoxide poor comets measured. The temporal variation in CO
abundance was seen to correlate with the rotation of the comet.
It should be remembered that A'Hearn {\it et al.} (2011) detected
strong CO$_2$ (CO$_2$/H$_{2}$O $\sim20\%$)
with the EPOXI spacecraft at a similar time.
Thus, Hartley 2 is CO$_2$ rich while CO poor.
\nocite{lufeweto2007,weaveretalHartley2011,ahepoxi2011}

Weaver {\it et al.} noted that CO/H$_{2}$O varies by a factor of 50
across all comets observed to date using UV data.
This is in agreement with IR observations (e.g. Paganini {\it et al.} 2014b).
\nocite{paganinilovejoy2014}
Though the most CO-rich comets
are long period comets, not all long period comets are high in CO (e.g.
C/2000~WM1 (LINEAR) is very low).  Even though CO can be produced as 
a daughter product of species such as CO$_2$ or H$_2$CO, 
Weaver {\it et al.} concluded that the HST
high spatial resolution suggests that the CO observed by HST originates
from the nucleus.

The  Solar Wind ANistropies (SWAN) all-sky Ly $\alpha$ camera on the 
Solar and Heliospheric Observatory (SOHO)
mission has been used regularly to monitor comets since its launch
in 1995.  Hydrogen is a dissociation
product of many species, the most notable of these H$_{2}$O.
To date, SWAN has been used to observe over 60 comets.
These comets are either observed as routine interlopers in the all-sky
images or can be specifically targeted for monitoring.
With its orbit at the L1 Lagrangian point, SOHO with SWAN is able
to observe comets at very small heliocentric distances as well as north
and south of the ecliptic. 
When a comet is very near to the Sun, the lifetime of H$_{2}$O against
photodissociation is extremely short and water dissociates completely
within the collisional zone.  Examples of comets observed
close to the Sun are C/2002~V1 (NEAT),
C/2002~X5 (Kudo–Fujikawa), 2006~P1 (McNaught) and 96P/Machholz~1
(Combi {\it et al.} 2011).  The closest passage to the Sun by a comet
observed by SOHO came in fall 2013 when comet C/2012~S1 (ISON) broke
up within a few solar radii, hours before perihelion (Combi {\it et al.} 
2014).
\nocite{combietalSWAN2011,combietalISON2014}

While the SWAN observations tell us only about the water production rate
of comets and not of other species, they are extremely valuable for being
able to monitor the production rates of comets as they
approach and recede from the Sun.  Combi {\it et al.} (2011) pointed out
that there are two types of behaviors: comets whose water production
rate increases with a steep slope as they approach the Sun and
those with a more shallow slope.  However, they pointed out that all of
the steep-sloped comets are Jupiter Family comets of the type
that have typical perihelia near 1\,{\sc au}.  
Indeed, the only periodic comets in the moderate-slope group are
153P/Ikeya-Zhang and 96P/Machholz~1. Neither is a Jupiter Family comet.
Ikeya-Zhang has a $\sim350$ year period.
Machholz 1 is in a highly inclined (58$^\circ$) orbit.

\subsection{IR Observations}

The lack of a permanent dipole moment in many of the cometary parent
molecules means that these molecules cannot be observed in the optical or
radio;
they have strong vibrational transitions in the IR. 
Studies of these spectral regions were not possible
until the 1970s when IR detectors were introduced to
astronomy.  The first detectors were not very sensitive and were also
very small. 

\begin{figure}
\centering
\includegraphics[scale=0.7]{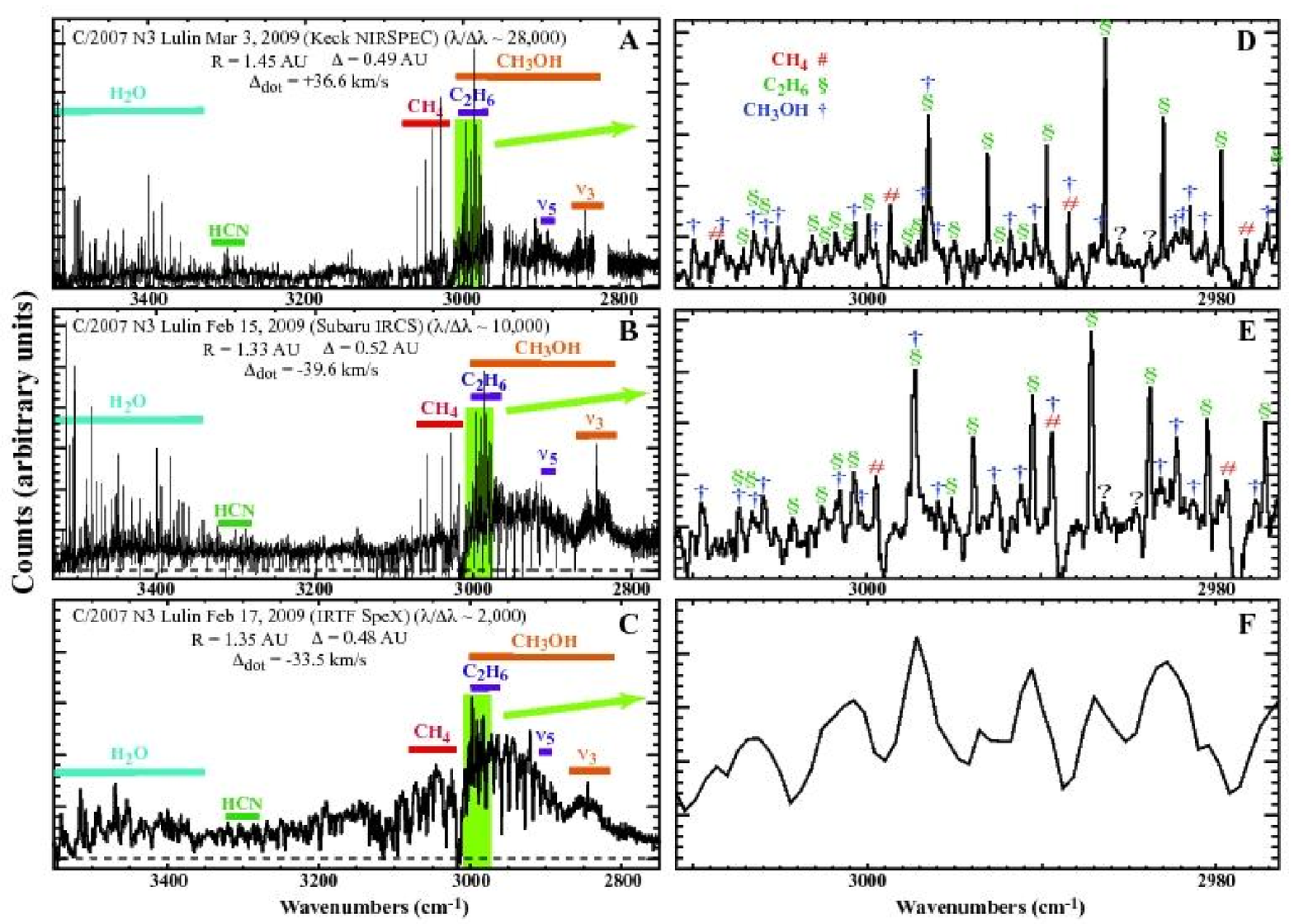}
\caption{Direct comparison of high and moderate-resolution spectra for C/2007 N3
Lulin.  Left panels show the complete spectra (before telluric correction)
from $\sim2.83-­3.64$ $\mu$m ($\sim 3550­-2750$ cm$^{-1}$). 
Right panels show a
zoomed-in portion ($\sim 3.32­-3.36$ $\mu$m) of these full spectra, as
indicated by green boxes in the left figures.  This illustrates two
important points: (1) High-resolution spectra are necessary for the
complete chemical interpretation of moderate-resolution spectra.  (2) IR
spectra are rich in cometary emissions underlining the importance of
spectral surveys.  Note that these spectra were obtained on different
dates so the geocentric Doppler-shift is different (particularly for the
NIRSPEC spectrum) so some lines are not present on all dates due to
telluric extinction. (Spectrum courtesey N. Dello Russo)}\label{lulin_ir}
\end{figure}

The IR offers some difficult challenges not seen in the optical part
of the spectrum.  The Earth's atmosphere has significant opacity
in the IR due to molecules such as CO$_2$ and H$_{2}$O.
Thus, there are broad regions of the atmosphere that
are opaque to the cometary photons.  This means that we can
only study molecules that occur in ``clean" regions of the spectrum.
It also argues for higher spectral resolution to avoid bad
regions of the spectrum.
Even the clean regions have significant, and variable, opacity so the atmosphere
must be carefully measured, with equal amounts of observation
time on sky and comet, and the atmosphere must be modeled and removed.
Figure~\ref{lulin_ir} demonstrates the importance of high
spectral resolution for IR work.

Most of the IR observations are in the 1--5$\mu$m region,
with most of the interest in the 2.5--5$\mu$m
bandpass. IR observations are typically limited by the regions
in the Earth's atmosphere that are most transparent. Doppler shifts of cometary
lines can move them from the good region into the more opaque regions of
the bandpass.  Limitations on time available to observe and the
spectral grasp can lead to different molecules being
detected for different comets.  

A variety of species in the IR spectra of comets have been detected, including
H$_{2}$O, CH$_{4}$, C$_2$H$_2$, C$_2$H$_6$, H$_2$CO, NH$_{3}$, 
CH$_3$OH, HCN, etc. All of these are seen in molecular clouds.
However, there are still some species seen in molecular clouds and not comets.
In addition to the parent species seen in the IR, fragments
such as OH, NH$_{2}$ and CN are observed.
Dello~Russo {\it et al.} (2013) include a high-resolution atlas of comet
103P/Hartley~2.  \nocite{dellorussoetal2013,muetal2001b}
The first such infrared spectra of a comet obtained with a cross-dispersed
echelle spectrometer appeared in Mumma {\it et al.} (2001) for
comet C/1999~H1 (Lee). A complete spectral atlas for comet Lee appeared
in Dello Russo {\it et al.} (2006). \nocite{dellorussoetal2006}

Determining production rates from IR spectra requires taking into account the
various excitation processes.  These are discussed
in detail in Bockel\'{e}e-Morvan {\it et al.} (2004).
As with the optical models, differences in model parameters
can make intercomparing results a bit problematic.
In recent years, intense activity has led to greatly improved
fluorescence models for primary volatiles in comets and improved
transmittance models for the terrestrial atmosphere (e.g. DiSanti
{\it et al.}, 2013; Gibb {\it et al.}, 2013; 
Kawakita and Mumma, 2011; Lippi {\it et al.}, 2013;
Radeva {\it et al.}, 2011; Villanueva {\it et al.}, 2012).
\nocite{gibbetal2013,villanueva2012b,villanueva2013b}
\nocite{bocrmuwe2004,disantigz2013,kamu11,lippietal2013,radevaetal2011,
villanuevaetal2012a}

The sample of comets observed in the IR is much smaller than the sample
in the optical because the IR observations require bigger telescopes
and the IR cannot be used to observe the fainter comets.
In addition, IR observations are generally obtained
over a smaller range of heliocentric distances than, for example optical
data, because the comets are fainter far from the Sun.
This is rapidly changing.  Paganini {\it et al.} (2013) detected
CO in the unusual comet, 29P/Schwassmann-Wachmann~1 at 6.26\,{\sc au}.
At this point,
about three dozen comets have been observed in some part of the IR.

IR observations are scattered over many papers
that concentrate on a single or a few comets, with brief mention
of groupings of comets. 
Since 2013, papers have been published on 
2P/Encke (Radeva {\it et al.}, 2013), 
21P/Giacobini-Zinner (DiSanti {\it et al.}, 2013), 
81P/Wild~2 (Dello-Russo {\it et al.}, 2014b), 
103P/Hartley~2 (Kawakita {\it et al.}, 2013; Dello Russo {\it et al.}, 2013,
Bonev {\it et al.}, 2013), 
C/2003 K4 (LINEAR) (Paganini {\it et al.}, 2015)
C/2009~P1 (Garradd) (DiSanti {\it et al.}, 2014),
C/2010`G2 (Hill) (Kawakita {\it et al.}, 2014), 
C/2012~F6 (Lemmon) (Paganini {\it et al.}, (2014a), 
C/2012~S1 (ISON) (Bonev {\it et al.}, 2014),
and 
C/2013~R1 (Lovejoy) (Paganini {\it et al.}, 2014b). 
Inspection
of this list shows that it is a mix of long-period and Jupiter Family
comets. However, it has been much harder to observe Jupiter
Family comets in the IR since they generally are at the faint end of
detectability.
\nocite{dellorussowild2014,paganinilovejoy2014,kawakitahill2014,disantigarradd2014,paganinilemmon2014,radevaencke2013,kawakitahartley2013,dellorussoetal2013,disantigz2013,bonevetal2014}
\nocite{bonevetal2013,paganinietal2015}

From the IR observations, there is an emerging picture that there are
three different kinds of comets: typical, organic enriched and organic
severely depleted (c.f. Mumma and Charnley 2011). 
However, as Mumma and Charnley pointed out,
taxonomies based on optical fragments and IR primary species are not
always in agreement.  As an example, 8P/Tuttle, a Halley Type Comet, is
found to be typical in C$_{2}$/CN in the optical but peculiar in the IR
because C$_2$H$_2$ and HCN are severely depleted and C$_2$H$_6$ is low,
in contrast to CH$_3$OH that is enriched (Bonev {\it et al.},2008).
\nocite{much2011,bonevetal2008}

The limited windows through the atmosphere that ground-based IR observations
must use mean that important parents cannot be observed from the ground.
The most important is CO$_2$.  CO$_2$ is important because it
represents a significant fraction of the ice (around 20\% is not
uncommon).  CO$_2$ is also much more volatile than H$_{2}$O. Thus, while the
activity of comets is controlled by the sublimation of water inside
$\sim3\,${\sc au}, outside of this region, it is CO$_2$ or CO that
control the activity (CO can either be a daughter or a parent).
Observations obtained with the AKARI spacecraft demonstrate this point,
showing that the CO$_2$ abundance relative to H$_{2}$O is much higher
outside of 2.5\,{\sc au} (Ootsubo {\it et al.} 2012).
\nocite{ootetal2012,rekeva13}
Some comets were observed both outside
and within $\sim2.5$\,{\sc au} and the driver of activity is shown to change.
No evidence was seen for a difference between Jupiter Family and long-period
comets.

Reach {\it et al.}. (2013) used the Spitzer Space Telescope to determine
the amount of CO$_2$ relative to H$_{2}$O.  While many of the Spitzer-observed
comets look to have abundances similar to the AKARI comets, Reach
{\it et al.} found a number of comets that were CO$_2$ poor.  Those
CO$_2$ poor comets are a mix of optically classified typical and depleted
comets. 
Reach {\it et al.} concluded that more of the CO$_2$ poor comets were
depleted than typical comets but they did not have taxonomic
classifications on several of the CO$_2$ poor comets. With recent
photometric observations, it turns out that the CO$_2$ poor comet group
contains 3 depleted comets, 5 typical comets and 1 unclassifiable
comet (Schleicher 2015, personal communication).
It should be noted that while the AKARI observations
are spectral, the Spitzer observations were imaging observations.
The filter that contains the CO$_2$ also contains a considerable
quantity of dust that was measured with a nearby continuum filter.  
Depending on the dust properties, it is possible that some of the
noted depletions are instead dust contamination.

While the AKARI and Spitzer spacecraft have been used to look at a large
number of comets, these spacecraft are available for observations for only
a limited amount of time. McKay {\it et al.} (2012, 2013) and
Decock {\it et al.} (2013, 2014) 
have used discussions of branching ratios for the processes
to produce [O~I] (e.g. Festou and Feldman, 1981) and 
have developed procedures and empirical release rates to use 
observations of the forbidden oxygen lines in the optical to {\it infer}
the abundance of CO or CO$_2$.   This is done by looking at
the ratio of the O($^1$S) (or green) line to the sum of the O($^1$D) (or red)
pair of lines.  Unfortunately, this method relies on lifetimes and branching
ratios that are not well determined in the lab and calibration against
the IR is sorely needed.
\nocite{fefe81}

A key use of IR spectra has been to map the ortho/para ratio of the gas
for a variety of molecules. If a molecule has H atoms that are
located symmetrically, then these molecules may have states that
have different spin orientations of the hydrogen.  As an example, for H$_{2}$O,
the two H atoms can have their spin orientation the same or
opposite one another. The first of these orientations is known
as the ``ortho" state, while the second is the ``para" state.  
A unique feature of the spin states is that it is not possible
to flip the state by radiative or collisional processes. Thus,
the ortho/para ratio (OPR) of the coma gas measures the ortho/para ratio of
the gas reservoir when the gas was frozen into the nuclear ices.
Measurement of the rotational distribution in each spin species can lead
to a measurement of the ortho/para ratio and thus to the nuclear spin
temperature.  Since most
of these molecules reach equilibrium at some temperature, the ortho/para
ratio measures the spin temperature of the ice (but see below).
Bockel{\'e}e-Morvan {\it et al.} (2004) discuss how these conversions
are achieved.

Mumma {\it et al.} (1987) were the first to measure an OPR in H$_{2}$O for
a comet.  Since then, about a dozen OPRs in water have been measured.
Mumma and Charnley (2011) summarized the values through 2011.
Recently, Bonev {\it et al.} (2013) and Kawakita {\it et al.} (2013) used this
technique to study the spin temperature in H$_{2}$O  for comet 103P/Hartley~2.
As discussed in these papers, they found a range of ratios that did not
always agree within the error bars.  However, in general, it appeared
that T$_{spin}$ was below 40\,K.
\nocite{bonevetal2013,kawakitahartley2013,shinnakaetal2011,muwela87,much2011}

Shinnaka {\it et al.} (2011) presented OPRs in NH$_{2}$ (optical) and 
NH$_{3}$ (IR) for
15 comets.  They found the same value for most comets, with SW3
standing out as being different.  Occasionally, OPRs can be measured
in other species (e.g. CH$_{4}$, c.f. Mumma and Charnley 2011).
These observations, however, are rarer because they
rely on being able to observe the comets at critical Doppler shifts.
In eight comets, spin temperatures for NH$_{3}$ and H$_{2}$O were relaxed 
and in agreement; spin temperatures for methane also agreed when co-measured
with H$_{2}$O or NH$_{3}$ (c.g. Mumma and Charnley, 2011).

The use of  OPRs was believed to be a sensitive measure of the
formational temperature of the ice.  However, recent laboratory
experiments (Fillion {\it et al.} 2012) pointed out that the speed
of desorption has a major affect on this ratio.
Thus, it remains to be seen how useful OPRs are as constraints on
temperature in the solar nebula vs. as taxonomic tools.
\nocite{fillionetal2012}

\subsection{Radio Observations}
Pure rotational transitions for molecules with dipole
moments occur in the radio portion of
the spectrum.  These simpler transitions make it easier to 
detect complex volatiles. In addition, radio spectroscopy has very high
spectral resolution, allowing for accurate measurements of spectral line
shapes and Doppler shifts in the cometary coma.  Thus, with radio observations
it is possible to measure the line-of-sight outflow velocity of the coma.
Radio telescope beams tend to be very large. Thus, the radio observations
measure the total abundance of a species but have to assume the
outflow is symmetric.  Mapping may be obtained by shifting the
beam center,  though the additional positions are  generally limited in
number.  The mapping can be improved upon with interferometry 
(see subsection \ref{interferometry}).

The earliest radio observations were made of the daughter species, OH,
at 18cm.  Since 1973, there have been more than 50 comets observed
with the Nan{\c c}ay telescope (Crovisier {\it et al.} 2002;
Colom {\it et al.} 2011).
\nocite{crcogebobo02,colom2011}

There are more than a dozen species seen in the radio in comets.
Species such as CO, H$_2$CO, CH$_3$OH, CH$_3$CN, HC$_3$N, HCN, NH$_{3}$,
HCOOH, HNCO, and H$_2$S are thought to be primary (parent) volatiles.
OH, CS, and HNC are likely to be product species.  As with the IR,
all of these species cannot be observed simultaneously so we
have detections of these species in some comets but not
others.  Thus, though more than 40 comets have been observed
in the radio, very few species have been observed in more than 10
comets (Figure~1 of Crovisier {\it et al.} 2009 gives an indication
of the number of molecules seen per comet for different types of
comets).
\nocite{crovisieretal2009}

The comet with the record for the most species observed is C/1995~O1
(Hale-Bopp).
This comet's brightness allowed for the detection of molecules that had never
before (or since) been seen and measurement of the behavior of the species
with heliocentric distance (Biver {\it et al.} 1997, 2002; Bockel{\'e}e-Morvan
{\it et al.} 2000). In addition to the molecules mentioned above,
SO$_2$, H$_2$CS, NH$_2$CHO, CH$_3$CHO, and HCOOCH$_3$ were detected.
The most complex molecule detected in the spectrum of Hale-Bopp
was {HO-CH$_2$CH$_2$-OH} (ethylene glycol) (Crovisier {\it et al.}. 2004).
\nocite{BiveretalHB1997,boetalHB2000,crovisieretal2004,BiveretalHB2002}

Hale-Bopp was an exceptionally bright comet, with a very high water production
rate.
Indeed, it was the brightest comet since the radio and IR techniques
have become mature and routine.
Therefore, it was possible not only to observe more molecules in Hale-Bopp
than in other comets, but also to observe them over a larger range
of heliocentric distances.  Most species were observed out
to distances of 3--4\,{\sc au}. However, HCN and CH$_3$OH were both
observed to 6\,{\sc au} and CO was observed at 14\,{\sc au}, some four
years after perihelion (Biver {\it et al.} 2002). Figure~5 of
Biver {\it et al.} shows the trends of the production rates of nine molecular
species as a function of heliocentric distance, both pre- and post-perihelion.
It is obvious from this figure that at large heliocentric distances
the CO is controlling the cometary activity, while within about
3\,{\sc au} the H$_{2}$O sublimation is driving the activity.

\begin{figure}
\includegraphics[width=\textwidth]{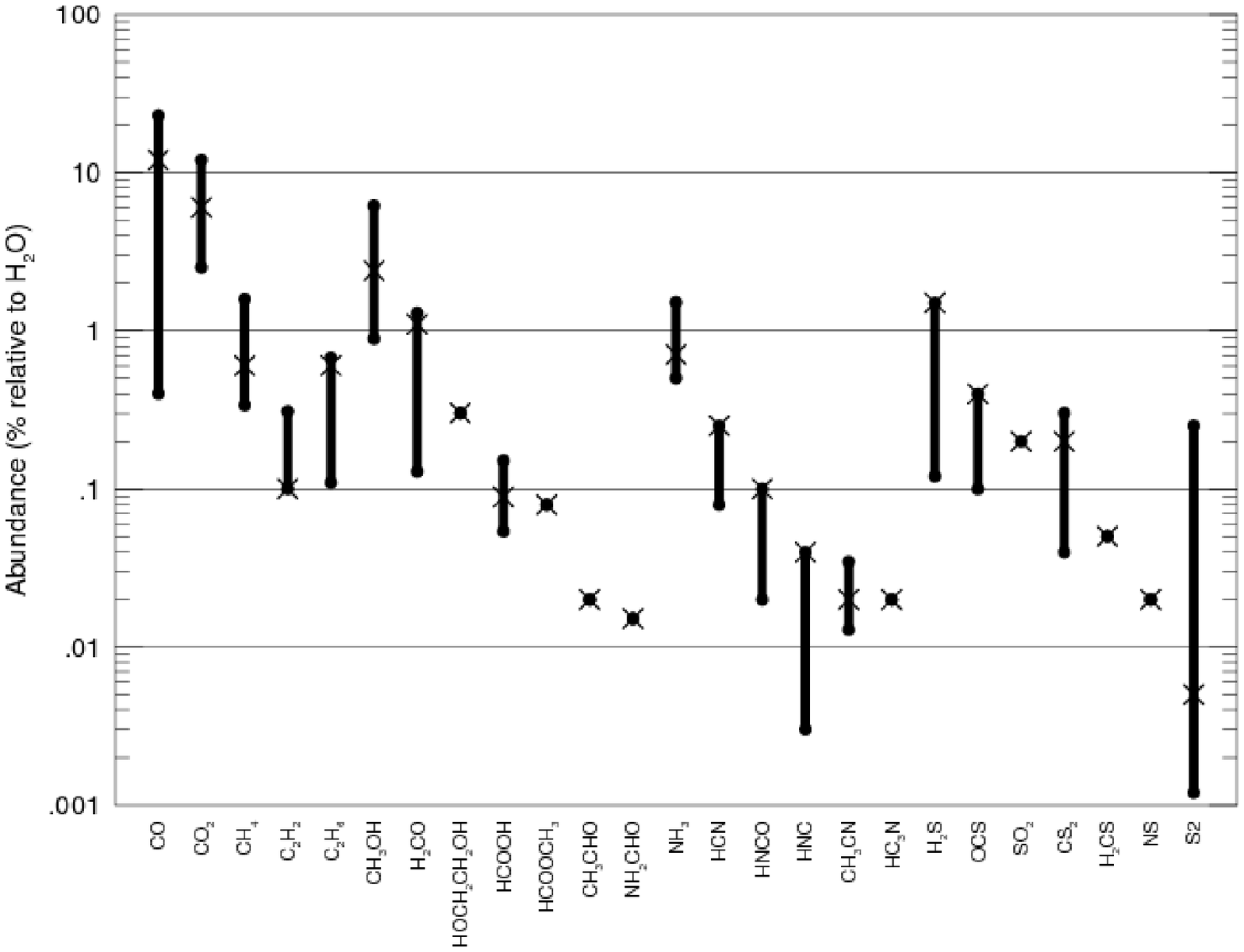}
\caption{The abundances for many of the IR- and radio-observed
species are shown as a percentage relative to water.  The bar
shows the range of values observed for each species.  
The "$\times$" marks the value determined for comet C/1995~O1 (Hale-Bopp).
CO, CH$_3$OH,
H$_2$CO, HCN, H$_2$S and CS$_2$ were all observed in at least
10 comets; all other species where observed in fewer.  The species
with a single value (an X-dot) were only observed in Hale-Bopp (based on
Figure~1 of Bockel{\'e}e-Morvan 2011).}
\label{abund}
\end{figure}

Crovisier {\it et al.} (2009) explored the radio-derived abundances of
several species relative to water for about 30 comets.  They found
no evidence for the three categories of typical, depleted and enhanced
comets seen in the IR, nor did they detect a difference between
Jupiter Family and other dynamical types of comets. They
noted that there is a wide spread of abundances relative to water
for most species. The exception to this is for their
observations of HCN, which, according to Crovisier {\it et al.},
show a narrow range of values.  
Figure~\ref{abund} shows the range of abundance values, relative to
water, found in the IR and radio for a number of comets.

\nocite{BoMo2011}

\subsubsection{The Rise of Radio Interferometry}\label{interferometry}

Combining the benefits of high spectral resolution with simultaneous spatial
mapping, radio interferometry is a powerful technique for probing the
distribution and kinematics of cometary gas and dust (eg. Blake et al. 1999;
Boissier et al. 2014).
Currently in Cycle 2 Early Science mode, the Atacama Large Millimeter/submillimeter
Array (ALMA) is a state-of-the-art radio/sub-mm interferometer under
construction at 5~km altitude in Chile's Atacama Desert.  Once completed in
2015-2016, using 66 antennae separated on baselines up to 16~km, extremely
detailed mapping of molecular line and continuum emission from cometary comae
will be possible across the frequency range 84-950~GHz, with an angular
resolution up to $\sim0.005''$. Cordiner {\it et al.} (2014) first
demonstrated the power of ALMA for quantitative measurements of the
distributions of molecules and dust in the inner comae of typical bright comets,
with spectrally and spatially-resolved maps of HCN, HNC, H$_2$CO and
0.9~mm dust continuum in C/2012 F6 (Lemmon) and C/2012 S1 (ISON).
\nocite{blakeetal999,boetal2014}

Observations of these two Oort-Cloud comets were made using the ALMA Band 7
receivers, covering frequencies between 338.6 and 364.6~GHz (0.82-0.89~mm)
using 28-30 12-m antennae (with baselines 15-2700~m, which provided an angular
resolution of approximately $0.5''$). Comet Lemmon was observed post-perihelion
on 2013 June 1-2 at heliocentric distance $r_H=1.75$~AU and ISON was observed
pre-perihelion on 2013 November 15-17 at $r_H=0.54$~AU.  The spectral
resolution was about 0.42~\kms. For further details see Cordiner {\it et al.} (2014).

\begin{figure}
\centering
\includegraphics[scale=0.32]{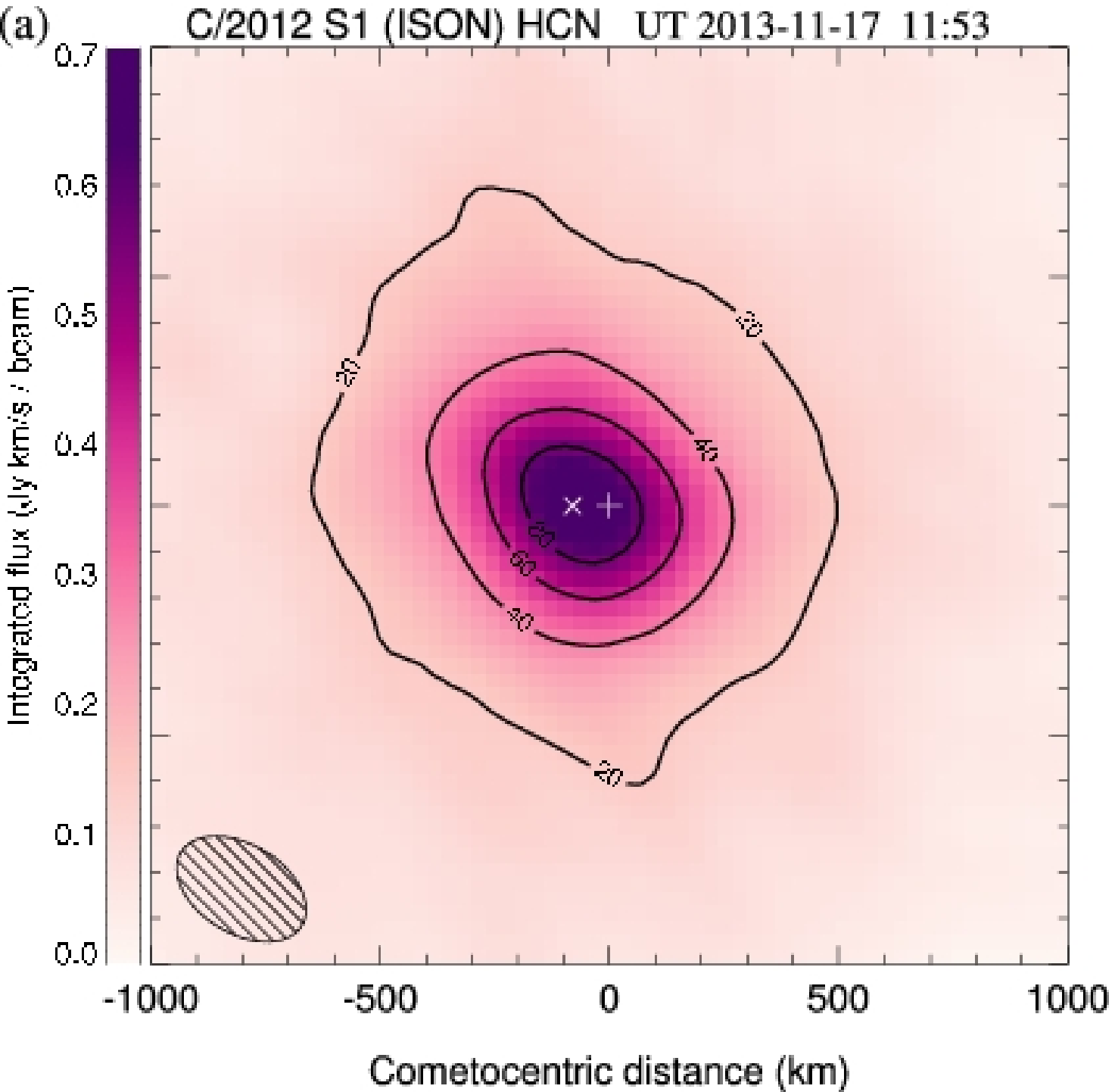}
\includegraphics[scale=0.32]{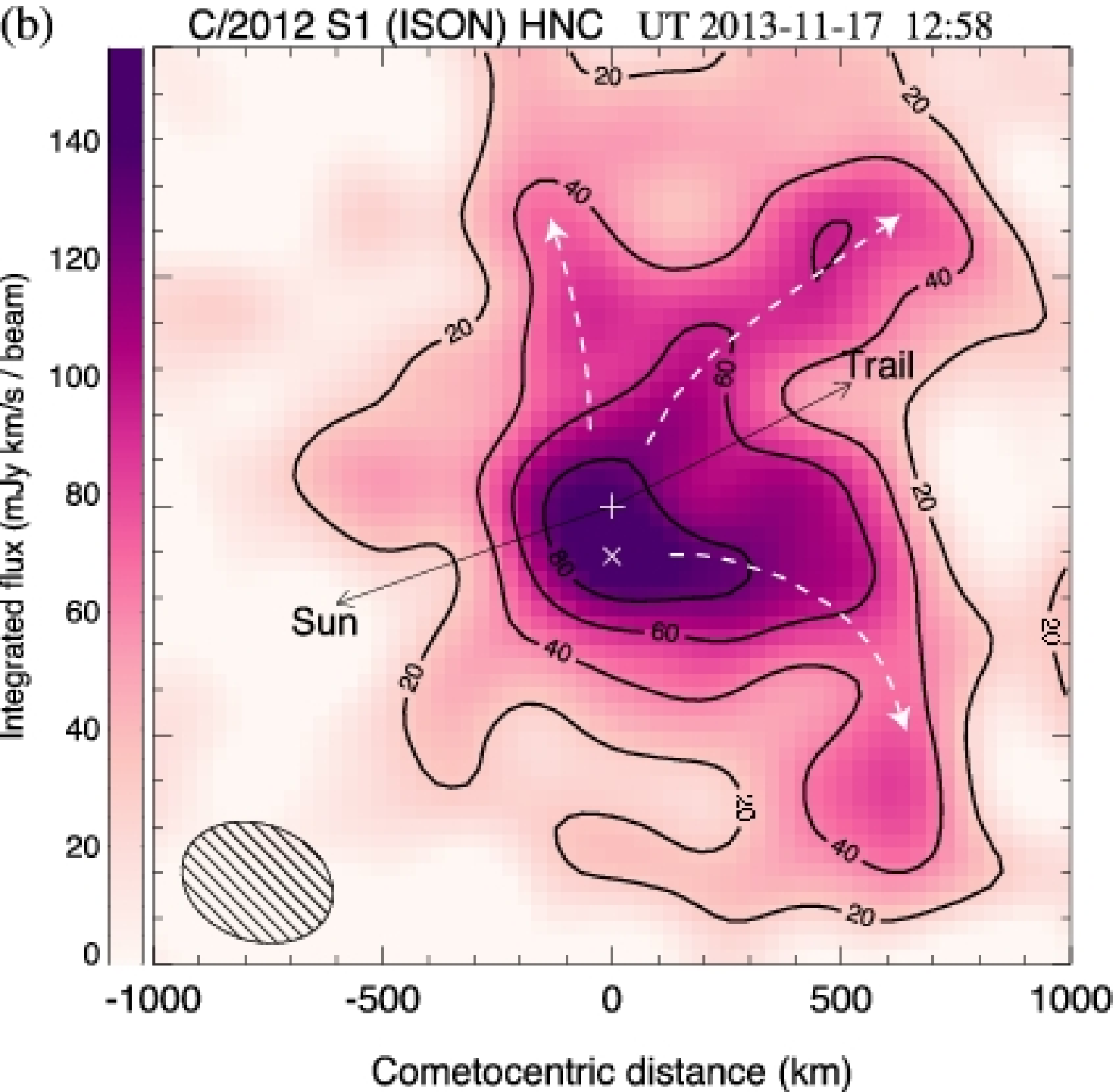}
\includegraphics[scale=0.32]{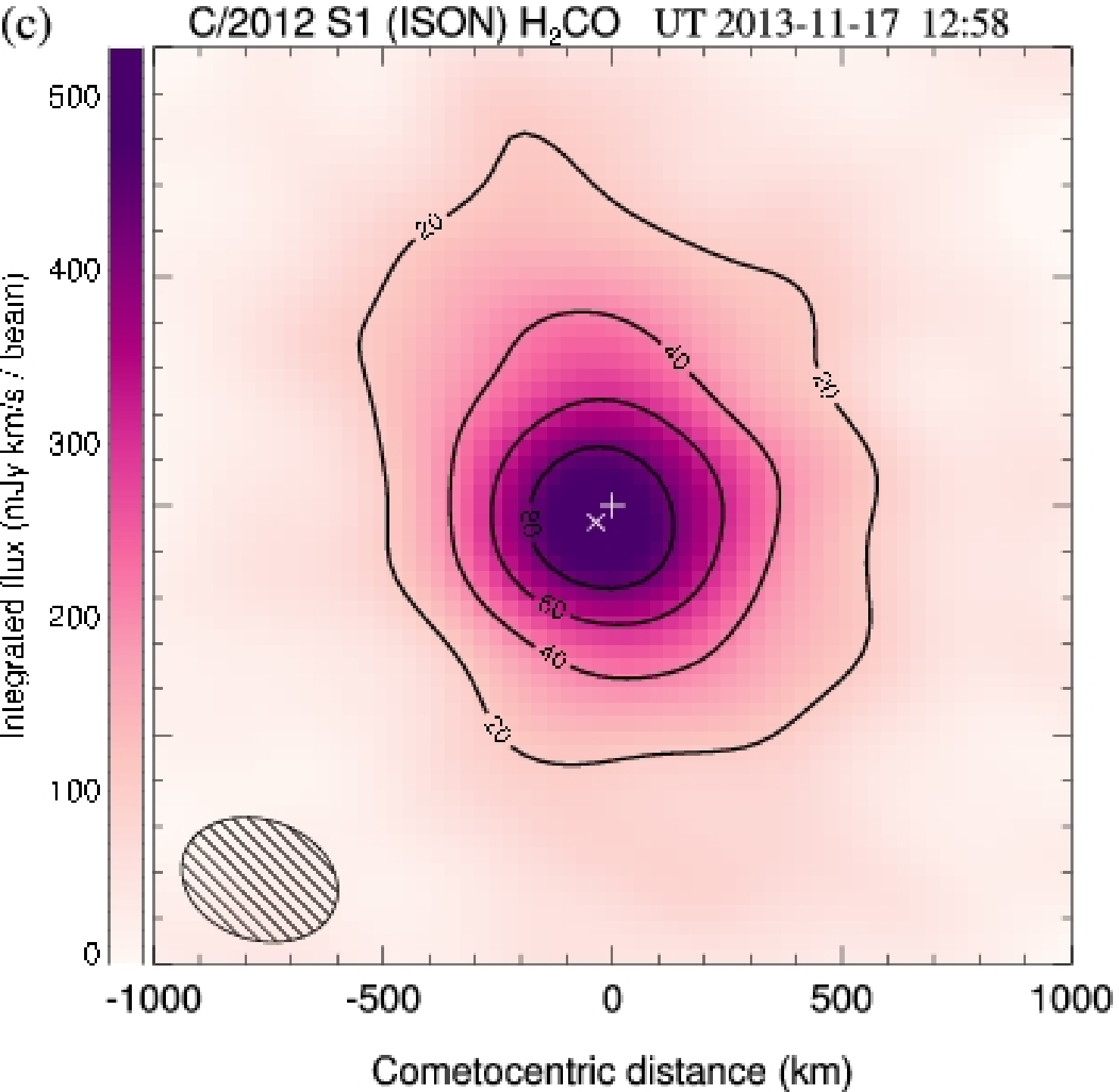}\\
\includegraphics[scale=0.32]{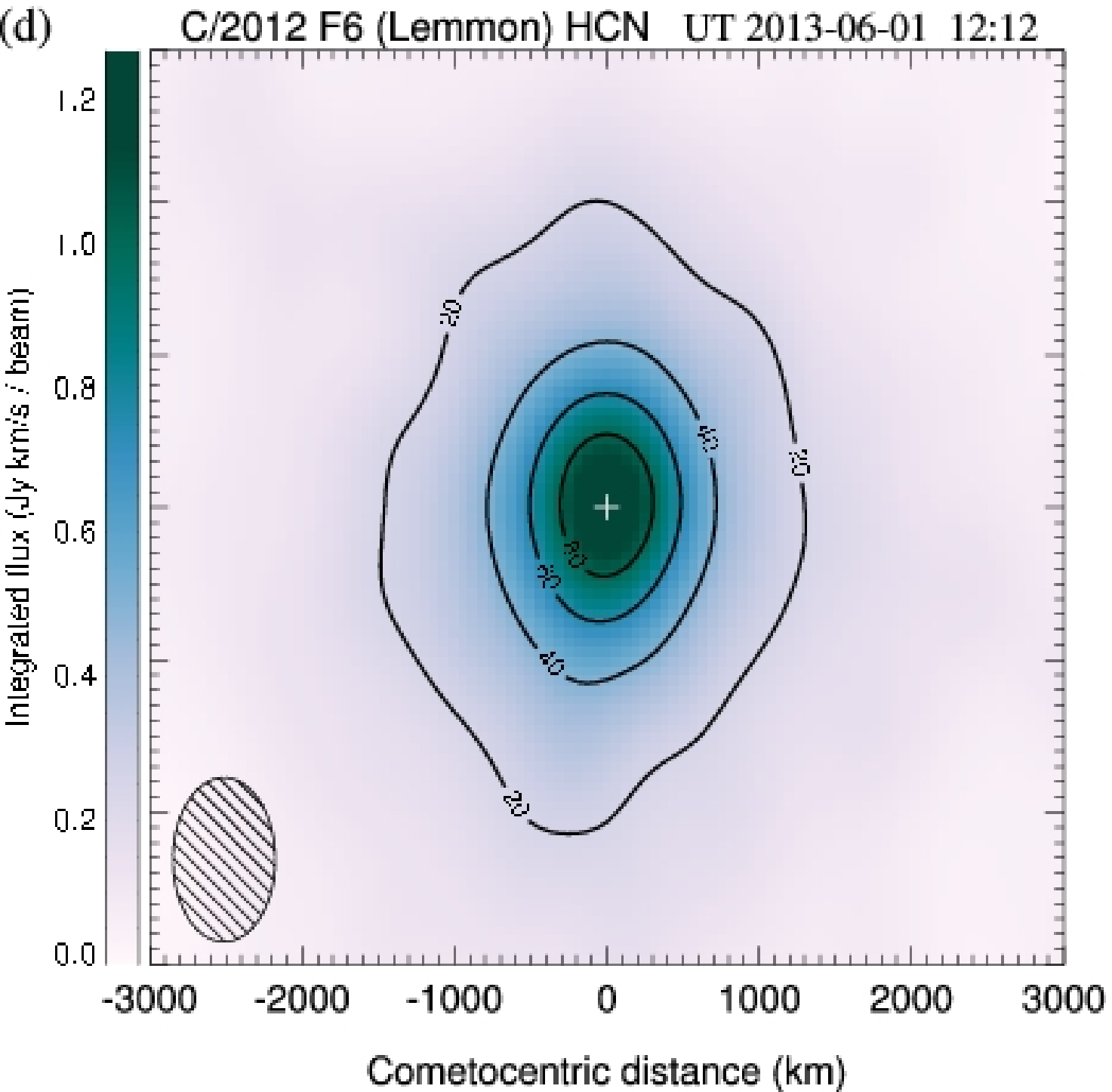}
\includegraphics[scale=0.32]{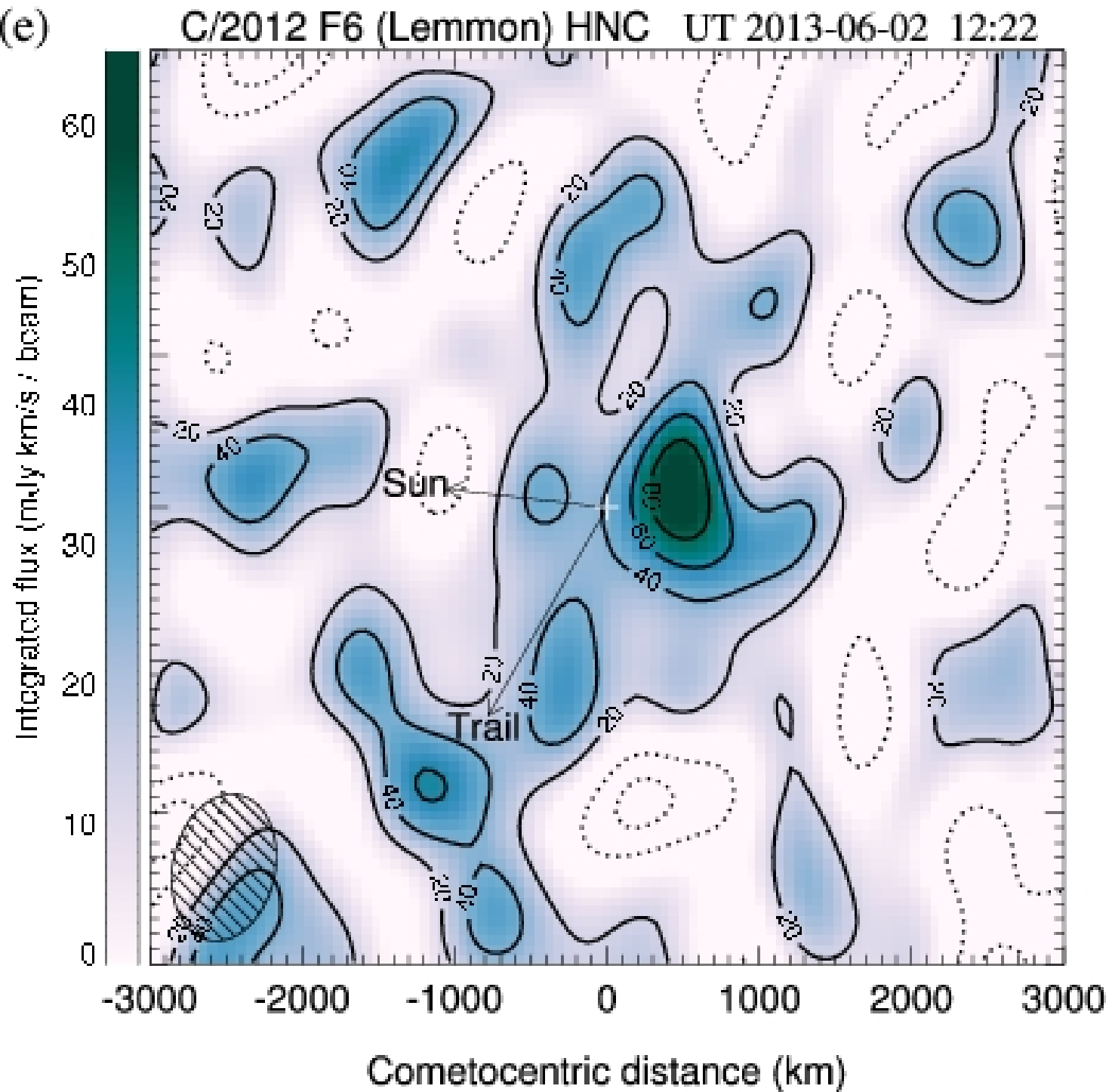}
\includegraphics[scale=0.32]{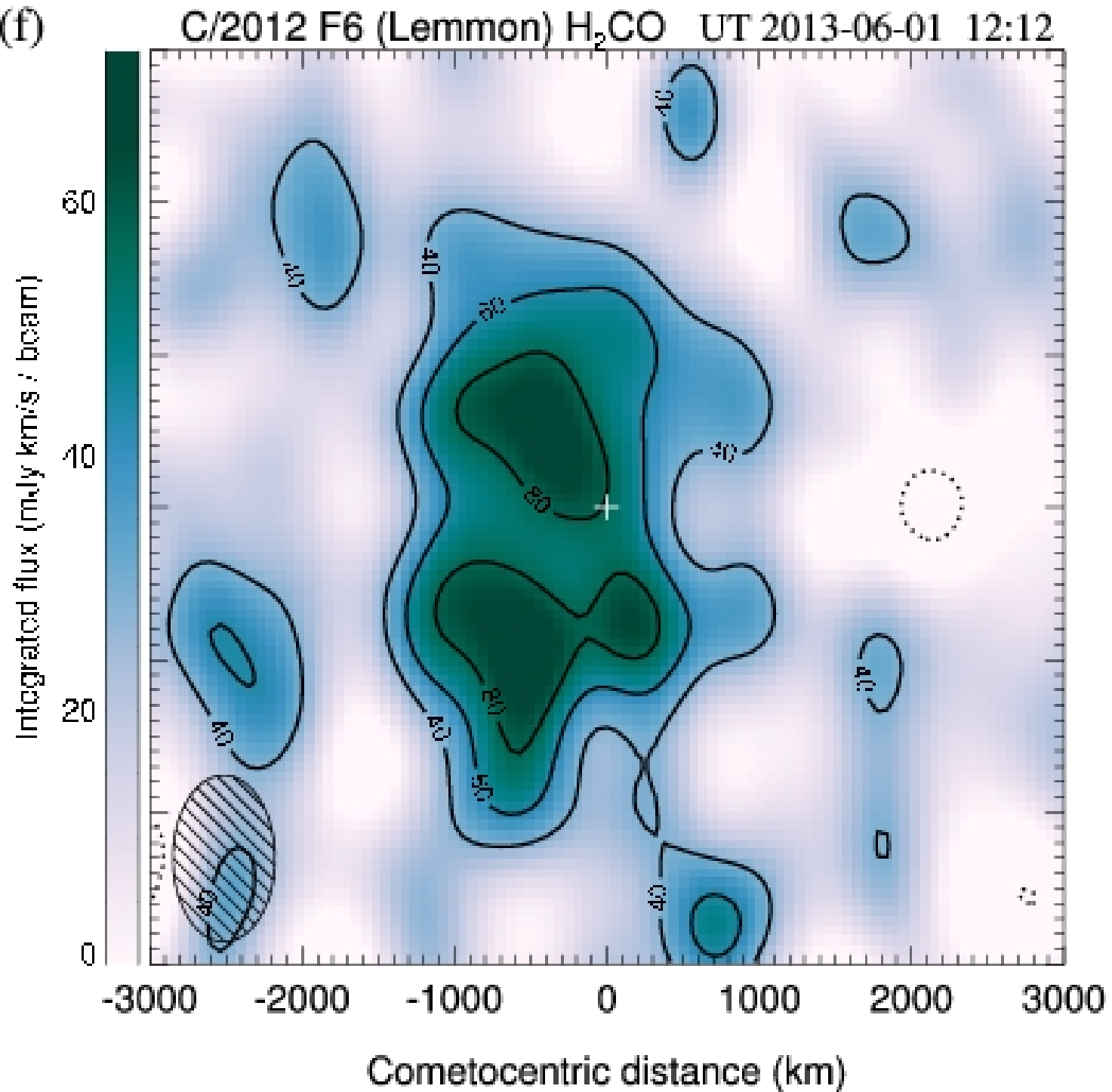}
\caption{Contour maps of spectrally-integrated molecular line flux observed
in comets S1/ISON (top row) and F6/Lemmon (bottom row). Contour intervals in
each map are 20\% of the peak flux. The 20\% contour has been omitted from
panel (f) for clarity. Negative contours are dashed. On panel (b), white
dashed arrows indicate putative HNC streams/jets. The RMS noise ($\sigma$,
in units of mJy\,beam$^{-1}$\,km\,s$^{-1}$) and contour spacings ($\delta$)
in each panel are as follows: (a) $\sigma=10.9$, $\delta=25.2\sigma$,
(b) $\sigma=13.6$, $\delta=2.3\sigma$, (c) $\sigma=11.0$, $\delta=9.6\sigma$,
(d) $\sigma=13.1$, $\delta=19.4\sigma$, (e) $\sigma=14.8$, $\delta=0.9\sigma$,
(f) $\sigma=13.7$, $\delta=1.1\sigma$. The peak position of the (simultaneously
observed) 0.9~mm continuum is indicated with a white `$+$'. For ISON, a white
`$\times$' indicates the integrated molecular emission peak position, offset
from the continuum (dust) peak in each case. Sizes (FWHM) and orientations of
the point-spread functions are indicated in lower-left (hatched ellipses);
observation dates and times are also given.
See Cordiner {\it et al.} (2014) for further details.}
\label{fig:maps}
\end{figure}

Figure~\ref{fig:maps} shows spectrally-integrated flux contour maps for the
observed molecules in each comet. Dramatic differences are evident between
different molecular species observed in the same comet, and between the same
species observed in the two comets. By eye, the HCN distributions in both
comets appear quite rotationally-symmetric about the central peak. For Lemmon,
no offset between the HCN and continuum peaks is distinguishable, whereas
ISON's HCN peak (indicated with a white `$+$') is offset 80~km eastward from
the continuum  peak (white `$\times$'). Shown in Fig. \ref{fig:maps}b, the HNC
map for ISON exhibits a wealth of remarkably extended spatial structure, with
at least three streams (identified at $>6\sigma$ confidence), emanating away
from the main peak (indicated with white dashed arrows). The majority of HNC
emission from both comets is asymmetric, originating predominantly in the 
anti-sunward hemispheres of their comae.

Formaldehyde also shows strikingly different distributions for comets Lemmon
and ISON (Figs. \ref{fig:maps}c and \ref{fig:maps}f),
highlighting the complex origin of
this species. Lemmon has a remarkably flat and extended H$_2$CO map, as
demonstrated by the size of the region traced by the 40\% contour compared
with the other maps. By contrast, the H$_2$CO distribution for comet ISON is
dominated by a strong, much more compact central peak, and has a relatively
symmetrical contour pattern, similar to that of HCN. 

Using the method of Boissier {\it et al.} (2007, 2014), Cordiner {\it et al.}
(2014) modeled the
interferometric visibility amplitudes for the observed species in order to
determine their origins within the coma. In both comets, HCN was found to
originate from (or within a hundred km of) the nucleus, with a spatial
distribution largely consistent with spherically-symmetric, uniform outflow.
By contrast, the HNC and H$_2$CO distributions were consistent with the release
of these species as coma products.  The H$_2$CO parent scale length was found
to be a few thousand km in Lemmon (at $r_H=1.75$~AU) and only a few hundred km
in ISON (at $r_H=0.54$~AU), consistent with destruction of the H$_2$CO
precursor by photolysis or thermal degradation at a rate which scales in
proportion to the solar radiation flux. The scale length for the putative
parent of HNC in comet ISON was found to be $\sim1000$~km.

The release of HNC and H$_2$CO as product species implies the existence of
organic precursor materials in the coma, which undergo sublimation,
photochemical and/or thermal degradation to produce the observed molecules in
the gas phase.  Heating or photolysis of materials such as 
grains, polymers or other macro-molecules, and their subsequent breakdown at
distances $\sim100$-10,000~km from the nucleus presents the most compelling
hypothesis for the origin of the observed H$_2$CO and HNC. The presence (and
composition) of the hypothesized macro-molecular precursors will be measured
by the COSIMA instrument on the Rosetta spacecraft during its encounter with
comet 67P/Churyumov-Gerasimenko in 2014-2015 (Kissel {\it et al.} 2007; Le~Roy
{\it et al.} 2012).

As ALMA's construction and commissioning are completed over the next two years,
the array will grow larger and even more sensitive, enabling the distributions
and kinematics of individual coma gases to be resolved on smaller scales than
have ever been achieved before; detections of new species should also be
possible. Such observations are expected to yield further fundamental new
insights into our understanding of cometary compositions.
\nocite{boetal2007,boetal2014,cordineretal14,kietal2007,leroyetal12}

\subsection{Caveat: What About Comets Near the Sun?}
The taxonomic studies discussed above generally refer to comets
that are at 1\,{\sc au} or larger heliocentric distances. While we
have evidence for a change of the volatile species that controls activity 
at distances greater than $\sim$3\,{\sc au} (generally CO or CO$_2$) compared
with within $\sim$3\,{\sc au} (H$_{2}$O), we rarely get to study comets
within the orbit of the Earth.  Those that we do get to study are
generally outside the orbit of Venus.  And it is rarer still to observe
a comet outside 1\,{\sc au} and follow it inside the orbit of Mercury.

\begin{figure}
\includegraphics[scale=0.7]{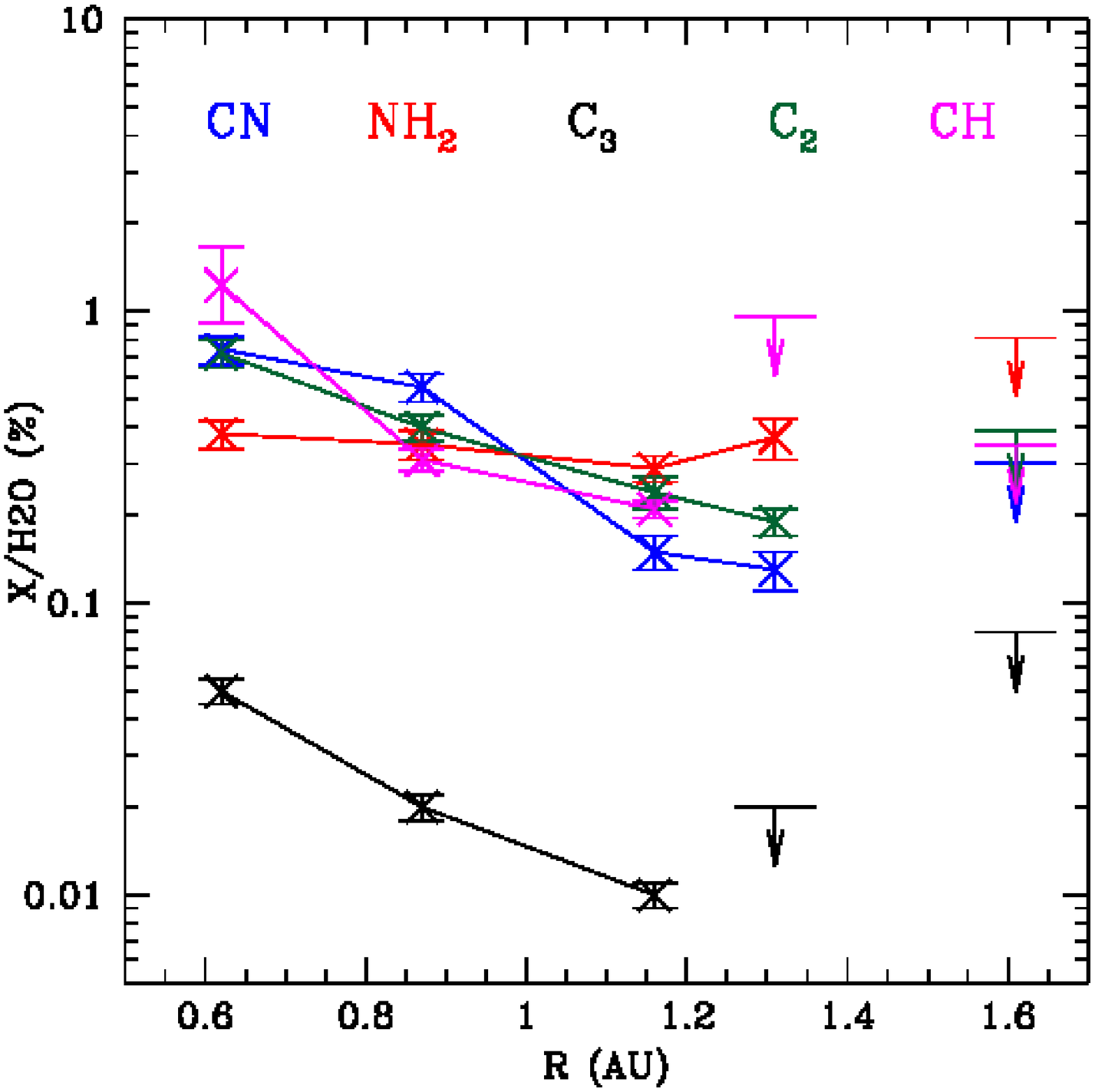}
\caption{The mixing ratios of various fragment species are shown as a
function of heliocentric distance for comet C/2012~S1 (ISON). The
data were obtained with various telescopes but all are high spectral
resolution observations.  NH$_{2}$ shows no change relative to H$_{2}$O
but the other species increased their fraction with respect to H$_{2}$O 
as the comet went from 1\,{\sc au} to 0.6\,{\sc au}.}\label{ISON}
\end{figure}

In fall 2013, we got just such an opportunity with comet C/2012~S1 (ISON). 
ISON was the very rare Sun-grazing comet that was discovered well in
advance of its perihelion passage (and subsequent destruction in this case).
At larger heliocentric distances, optical observations showed this
comet to be typical in some species, depleted in others and even
enhanced in some species.
IR observations of
C$_2$H$_2$ and HCN seemed typical; CH$_3$OH, C$_2$H$_6$ and CH$_4$
were depleted; NH$_{3}$ was enhanced (Dello Russo {\it et al.}. 2014a;
DiSanti {\it et al.} (2015).  Stranger still, DiSanti {\it et al.} 
noted that the H$_2$CO started out depleted, trended to typical and
ended up enriched as the comet went from $>1$\,{\sc au} to $<<1$\,{\sc au}.
\nocite{dellorussoACM2014,disantiISON2015,mckayACM2014,opitomACM2014}
Figure~\ref{ISON} shows what happened in optical spectra as a function of
time (McKay {\it et al.} 2014).  With the exception of NH$_{2}$, the
normal fragment molecules appear to have increased in production
relative to water as the comet moved well inside of 1\,{\sc au}.  
Opitom {\it et al.} (2014) saw similar behavior for C$_{2}$ and CN production
relative to OH using the TRAPPIST telescope monitoring the comet from
1.3 $\rightarrow$ 0.3\,{\sc au} inbound.
In addition, examination of the top row of the ALMA data
in Figure~\ref{fig:maps} shows
that there are large differences in the distribution of the different
species.

\subsection{Mass Spectrometer Measurements}
All of the above discussion was based on remote-sensing observations of
comets and includes observations of several hundred comets. Ideally,
one would like to take samples of the ices back to the laboratory
for measurement, especially if they could be kept
in the ice state so that the structure could be preserved and chemistry
prevented. However, sampling and transporting a sample back to
the lab is a process that is still on the horizon.  Alternatively,
one can take the laboratory to the comet.  The Rosetta mission is
doing just that, with samples of the coma from the main spacecraft
and samples of the ice from the Philae lander.

\begin{figure}
\centering
\includegraphics[scale=0.6]{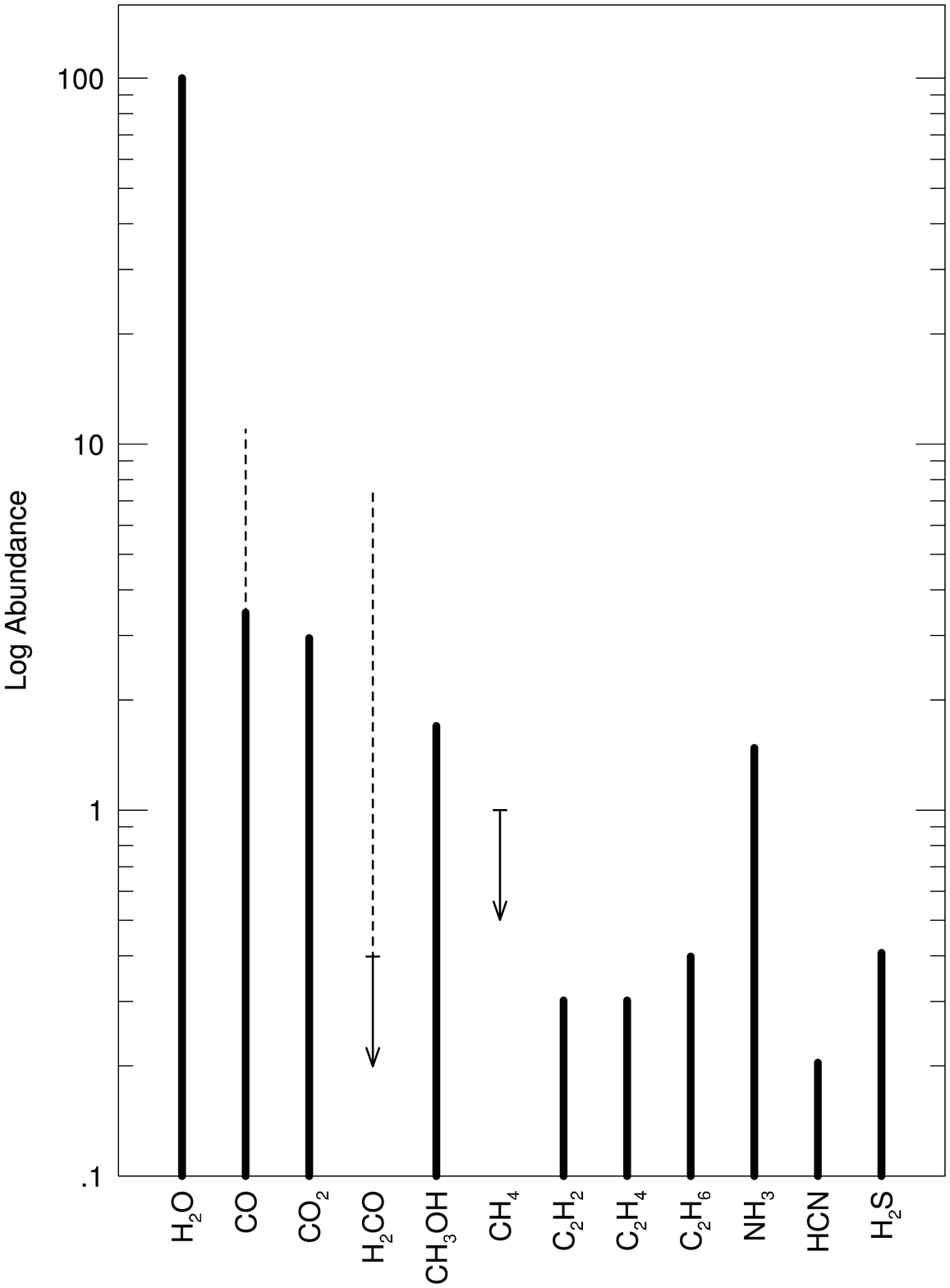}
\caption{The abundances of various species that were detected with
the mass spectrometers onboard the Giotto spacecraft are shown.
Heavy solid lines show the abundances for species detected in comet 1P/Halley.
Upper limits for the coma are shown with downward
arrows.  The lighter dashed lines show the additional contribution
of the extended sources in the coma. 
However, note that Giotto's closest approach to the nucleus was 1100\,km
so some species could be 
produced by chemistry in the first 1100 km of outflow, while others may be released entirely or in part from the nucleus
Based on Eberhardt (1999).}
\label{massspec}
\end{figure}

Prior to Rosetta, {\it in situ} direct measurements of the cometary material 
with a mass spectrometer have only
been
accomplished once, for comet 1P/Halley.  The Giotto spacecraft
was successful in sampling the coma during its flyby and analyzing
the composition.  The results of this have been discussed by
Eberhardt (1999).  \nocite{eb99}
It was the Giotto mass spectrometer that first measured many
primary volatiles leading to a detailed comparison of relative
abundances.
Figure~\ref{massspec} graphically shows the values, taken
from Table~1 of Eberhardt, for the 
abundances of the ten molecules detected as direct contributions
from the nucleus for comet Halley.  There are also two upper limits.
In addition, this experiment showed that some species,
CO and H$_2$CO,
that were directly produced at the nucleus, also resulted from
an extended source (however, note that the spacecraft was never closer to 
the nucleus than 110\,km so had no spatial information closer than that). 
CO obviously could be a parent as well as a daughter
of CO$_2$, but the H$_2$CO extended source is rarely seen in comets
(but see Subsection~\ref{interferometry} above).
However, as Eberhardt pointed out, the lack of detection of two
sources from the ground-based observations is more a function of the difficulty
of differentiating the sources at the spatial scale of the ground-based
observations.
The first measurements of the D/H ratio in
H$_2$O were obtained (with IMS and NMS on-board 
Giotto), providing D/H values of about $3 \times 10^{-4}$ (Balsiger {\it et al.},
1995; Eberhardt {\it et al.}, 1995).
\nocite{jess99,baalge95,ebrekrho95}

While these mass spectrometer observations of Halley are direct measurements, 
it should be remembered that they were still measuring the coma
gas after it had flowed off the nucleus.  Thus, some of the species
had already undergone some chemistry and photo processes. In addition,
the nature of a mass spectrometer means that there will be break-up
of some of the species going through the instrument.  Thus, to
interpret the measurements of the complex mass spectrum, one needs 
a detailed model of the {\it expected} abundances to match.
This is especially true of species that are 
available at the percent level or lower. In addition, depending
on the mass resolution of the instrument, there is a limit on
differentiating species of similar mass.  An important example
is CO vs. N$_2$, both at 28\,amu.  The Rosina spectrograph onboard
Rosetta has improved resolution from the Giotto instruments but
still relies on a model to interpret some of its results.

\subsection{Gas Composition: Rosetta Capabilities and Early Results}
There are instruments on both the Rosetta orbiter and Philae lander designed
to determine the composition of the coma.  These range from remote sensing
cameras with filters or spectrometers to coma-sniffing mass spectrometers.

On the orbiter, ROSINA (Rosetta Orbiter Spectrograph for Ion
and Neutral Analysis; Balsinger {\it et al.} 2007)
is a mass spectrometer with three parts: DFMS (Double
Focusing Mass Spectrometer, a high mass resolution spectrometer);
RTOF (Reflectron Time-of-Flight Mass Spectrometer, a high mass range
spectrometer) and COPS (pressure sensor to measure the total and ram
pressure).  DFMS can be used in restricted ranges to separate critical
species such as $^{13}$C vs. CH or CO vs. N$_2$.  RTOF can identify 
organic molecules, such as PAHs.  MIRO (Microwave Instrument for the
Rosetta Orbiter; Gulkis {\it et al.} 2007) is used to monitor H$_{2}$O,
CO and CO$_2$ abundances.
Alice (Stern {\it et al.} 2007), a UV imaging spectrometer, monitors coma
features such as the CO fourth-positive system, Ly $\alpha$, O, as well as 
scans the surface.  VIRTIS (Visible and Infrared Thermal Imaging 
Spectrometer, Coradini {\it et al.} 2007) is used to determine the
composition of the ices on the surface of the nucleus.
These instruments will be used over a wide range of heliocentric
distances as the spacecraft flies alongside the comet from rendezvous
through perihelion.

On the lander, samples are analyzed with COSAC (COmetary SAmpling and
Composition Experiment; Goesmann {\it et al.} 2007) and
Ptolemy (Wright {\it et al.} 2007).  COSAC is an evolved gas analyzer
that will concentrate on elemental and molecular composition.  Ptolemy
is an evolved gas analyzer that will ephasize measuring the isotopes.
The length of time that the lander can be used to sample the
gas depends upon power replenishment for the batteries.
On the original descent to the surface, the lander bounced
several times and was only able to send back a limited amount of data.
However, the lander has communicated with the ground again 
after the comet got closer to the Sun, starting in June 2015.

Early results from these instruments have shown a wealth of detail.
MIRO detected the ground-state rotation line of H$_{2}$O in June 2014,
when the comet was 3.9\,{\sc au} from the Sun (Gulkis {\it et al.}, 2015).
This was a blue-shifted asymmetric line, indicating that it came
from the day side of the comet.  MIRO observations showed that
the surface was very insulating.

Early ROSINA observations showed that the density of gas varied with
the rotation period and the latitude.  CO$_2$ and H$_{2}$O varied with
different rotational phases; the CO$_2$ and H$_{2}$O arise
from different places on the nucleus (H{\"a}ssig {\it et al.}, 2015).
The ratio between H$_{2}$O, CO and CO$_2$ varies quite substantially
along the spacecraft trajectory. ROSINA was also used to detect
various isotopes of H and O (Altwegg {\it et al.}, 2015)
The D/H was higher than other comets. ROSINA was able to detect both
CO and N$_2$, with modeling needed to derive abundances (Rubin
{\it et al.}, 2015).  They found that N$_2$/CO is severely depleted
relative to the protosolar values.

Together, the orbiter and lander observations provide ``ground-truth"
about comet 67P/Churyumov-Gerasimenko (CG).  We already know
from ground-based observations that CG is one of the
carbon-chain depleted comets. It will take all of the
observations of comets detailed in previous sections to place 
CG into context.

\nocite{2007SSRvBalsinger,2007SSRvGulkis,2007SSRvCoradini,2007SSRvStern,2007SSRvWright,2007SSRvGoesmann}
\nocite{rubinetalcg2015,altweggetalcg2015,hassigetalcg2015,gulkisetalcg2015}
\section{Solid Particle Composition and Properties}

This section summarizes our understanding of the composition of solid particles
in cometary comae. It is derived from flybys 
performed prior to the Rosetta rendezvous, from Earth (or Earth's orbit)
remote spectroscopic and polarimetric observations of
various comets, e.g., comet 1P/Halley, the first to experience a close 
spacecraft flyby,
comet C/1995 O1 (Hale-Bopp), a very active Oort 
cloud comet, and some rather bright Jupiter family comets (JFCs).
Solid particles are ejected with gases whenever the surface of a nucleus is 
close enough to the Sun to allow ice sublimation. Solid 
components in cometary comae appear to be mostly composed of refractory
materials such as silicates, organics and amorphous
carbon. Refractory dust particles present mixtures of different mineralogies.
Icy grains, which rapidly evaporate, and semi-refractory carbonaceous components
that suffer photolysis, may also be found in inner cometary comae. 
 
\subsection{Evidence from Cometary Flybys}
\subsubsection{1P/Halley flybys: First Discoveries on Cometary Dust Composition}
 
Major discoveries came from the dust mass spectrometers on board Vega 1 and 2
(PUMA) and Giotto (PIA) spacecraft in March
1986. The relative velocity of the probes and comet Halley on its retrograde
orbit was so high (about 70 km~s$^{-1}$) that only
information on the atomic composition could be obtained (Kissel {\it et al.},
1986a,b). From spectrometric analyses (mostly PUMA 1), 
dust particles were found to consist not only of a mixture of major
rock-forming elements (e.g., silicates, metals, sulphides, with
Mg, Si, Ca, Fe), but also, quite unexpectedly, of
carbon-hydrogen-oxygen-nitrogen compounds, the so-called ``CHONs". The
presence of large polymers of organic molecules was also suspected from the
positive ion cluster composition analyzer (with 
RPA on-board Giotto), the most refractory polymers being likely part of dust
particles (e.g., Krueger {\it et al.}, 1991), and from the
three-channel spectrometer (TKS) on board Vega 2 (e.g., Moreels {\it et al.}, 1994).
It was suggested that some gases detected in
the coma were degradation products of polymers of formaldehyde on cometary
solid particles (Cottin {\it et al.}, 2004).
\nocite{kietal86a,kietal86b,krkoki91,cogabera04,moclhebrro94}

The very low geometric albedo of the nucleus surface, of about 4\% (e.g.,
Keller {\it et al.}, 1986), was also a clue to a highly porous
surface, with its slightly reddish color suggesting the presence of
dark organic compounds. Moreover, local data on dust impacts (from the DID
instrument) and on light scattered by dust (from the 
OPE instrument) could be compared along the Giotto trajectory (Levasseur-Regourd
{\it et al.}, 1999). These two {\it in situ} data sets, with 
their similarities and correlations, offered a diagnostic test of the validity
of cometary coma models. Significant increases in local 
intensity and decreases in polarization for distances to the nucleus below 2000
km suggested the presence of small icy particles in 
the innermost coma.
(Levasseur-Regourd {\it et al.}, 1999). Fits of the data (in the 
2000-10000 km nucleus distance range) with a dust dynamical model indicate
that the grain size distribution index was of about 
-2.6, that the density of dust particles was very low, of about 100 kg~m$^{-3}$,
and that their geometric albedo was about 4\%, 
suggesting the dust particles to be both very fluffy and dark
(Fulle {\it et al.}, 2000). 
\nocite{kelleretal86,lemchafu99,fulemcha00}
Rocky and carbonaceous materials were mixed together on a very fine scale and
isotopic abundances appeared to be solar 
(Jessberger, 1999).

\subsubsection{81P/Wild~2 flyby: Dust Sample Collection}
Stardust collected about 4 mg of cometary material from comet 81P/Wild 2 in
January 2004, capturing dust particles ejected into the coma using aerogel, 
with a comet-spacecraft
relative speed of 6.1 km~s$^{-1}$. After delivery of the samples to Earth in
January 2006, continued analyses provided valuable insights into the
properties and composition of solid cometary material. Dust impact data
gathered (by DFMI) during the flyby indicated that dust particles were fragile
aggregates fragmenting as they were evolving in the inner coma of the comet
(Tuzzolino {\it et al.}, 2004). Chemical composition (from the CIDA dust mass
spectrometer) indicated the predominance of organic matter with light elements
present as gas phases, whereas the dust was rich in nitrogen-containing species,
with the presence of some sulfur ions (Kissel {\it et al.}, 2004). 
\nocite{tuz2004,kikrsicl04}

Analysis of the Stardust aerogels tracks showed the typical size range of dust particles
to be between 5 and 25 $\mu$m. These particles are a 
mixture of compact and cohesive grains (65\%) and friable less cohesive
aggregated structures (35\%) with constituent grains of a 
size less than 1 $\mu$m and a size distribution consistent with the ones
derived in the comae of comets (H{\"o}rz {\it et al.}, 2006). The 
collected particles are chemically heterogeneous at the largest scale
(of the order of 1 $\mu$m). The mean elemental composition 
suggests a CI-like composition consistent with a bulk solar system composition
for primitive material (e.g., Brownlee {\it et al.} 2012). The 
majority of the Stardust particles appear primarily composed of ferromagnesian
silicates with a larger range of Mg-Fe content than other comets
(Zolensky {\it et al.}, 2008), Fe-Ni sulphides and Fe-Ni metal. The 
range of olivine and low-Ca pyroxene compositions indicates a wide range of
formation conditions, reflecting a large scale mixing 
between the inner and outer protoplanetary disk (Zolensky {\it et al.}, 2006).
No hydrous phases of silicates were detected, which suggests a lack
of aqueous 
processing of Wild~2 dust (Keller {\it et al.}, 2006; Zolensky {\it et al.}, 2011).
Mg-carbonates were detected in Stardust samples, and they may be
produced by large KBO collisions or by nebular condensates (Flynn {\it et al.}, 2008).
\nocite{flyetal2008,zolenskyetal2008,zofrle2011}
The material accreted includes
Al-rich and Si-rich chondrule fragments together with 
some CAI-like fragments. These materials, combined with fine-grained components
in the tracks, are analogous to components in 
unequilibrated chondrite meteorites and clustered interplanetary dust particles
collected in Earth's stratosphere, so-called ``IDPs" (Joswiak {\it et al.}, 2012). 
\nocite{horzetal06,brjoma12,zolenetal2006,kelleretal06,joetal12}

Organics found in comet 81P/Wild~2 dust samples show a heterogeneous and
unequilibrated distribution in abundance and 
composition. Their characterization suggests that amorphous carbon and organic
carbon are dominant (Matrajt {\it et al.}, 2008). 
The carbon isotopic composition of glycine strongly favors an non-terrestrial
origin for this compound (i.e. not contamination), making it the first
amino acid 
detected in cometary material (Elsila {\it et al.}, 2009). 
\nocite{elgldw09,matrajt08}

H, C, N and O isotopic compositions are heterogeneous in the samples, but
extreme anomalies are rare, indicating that Wild~2 is 
not a pristine aggregate of presolar materials. Few grains with $^{17}$O and
$^{16}$O isotopic anomalies were found, showing them to be a 
minor phase of the comet, and indicating the presence of both presolar material
and material formed at high temperature in the 
inner solar system and transported to the Kuiper belt before comet accretion
(McKeegan {\it et al.}, 2006). The presence of deuterium 
and $^{15}$N excesses also suggests that some organics have an
interstellar/protostellar heritage (Sandford {\it et al.}, 2006). The oxygen 
isotopic compositions are consistent with chondritic material trends and the
presence of one additional $^{16}$O enriched reservoir 
(Nakashima {\it et al.}, 2012). The $^{26}$Al-$^{26}$Mg isotope presents no evidence
of radiogenic $^{26}$Mg, thus indicating that part of the 
cometary grains have formed more than 1.7 million years after the oldest solids
in the solar system, the calcium- and aluminum-rich 
inclusions (CAIs). This material must have been incorporated into the comet
several million years after the CAIs formed (Matzel {\it et al.} 2010). 
\nocite{McKeegan2006,sandfordetal06,nak2012,maetal2010}
One particular Wild~2 chondrule-like fragment, called Iris, is similar
to CR chondrite materials, and was age-dated to more than 1.8 million
years after CAI formation (Ogliore {\it et al.}, 2012).  Ogliore {\it et al.}
concluded that comet
formation probably occurred after Jupiter formation and about 2 million
years after CAIs.
\nocite{oglioreetal2012}

It nevertheless needs to be noted that, due to the fragility of the dust
particles and the relatively high speed of collection by Stardust, the analyses of
the composition of the particles and their interpretation is perturbed by
the mode of collection, which induced partial melting of some of the material
(e.g., preferential loss of light elements combined with possible
amorphization and
graphitization, Spencer and Zare, 2007; Fries {\it et al.}, 2009).
\nocite{spza2007,frbukest09}

\subsubsection{9P/Tempel~1 and 103P/Hartley 2: Icy Grains and Aggregates}
The nucleus of comet 9P/Tempel~1 was hit in July 2005 by a projectile released
by the Deep Impact spacecraft  (A'Hearn {\it et al.}, 
2005). It allowed comparison of the properties of materials found in the coma,
possibly originating from a processed surface, to sub-surface materials
released by the impact.
After the impact, there was an enhanced UV scattering that persisted for about 
20 minutes.
Strong absorptions in IR spectra near 3 $\mu$m were noticed,
providing evidence for icy grains in the plume
(Schulz {\it et al.}, 2006; Sunshine {\it et al.}, 2007). 
\nocite{ahdeepimpact,sunshine2007,schulz2006}
Before impact, the C$_2$H$_6$/H$_{2}$O ratio in the coma was depleted by a 
factor of three relative to organics-normal comets but after impact it 
increased to the organics-normal value (~0.6
hypervolatile ethane was more depleted in the surface layer than in the 
excavated material (Mumma et al. 2005).
\nocite{mudeepimpact2005}

The Deep Impact spacecraft was retargeted to comet 103P/Hartley~2, which it
flew by in November 2010 as part of the EPOXI 
mission. The nucleus was discovered to be bi-lobed in shape, with knobby
terrains on the lobes and relatively smooth regions on 
the waist, to have an average geometric albedo of about 4\%, and to present a
clustering of jets on parts of the lobes (A'Hearn et 
al., 2011). At closest approach, many images suggested the presence of large
chunks of icy particles, with sizes possibly reaching 
tens of centimeters. Further analyses confirmed the existence of such
aggregates, sublimating on their sunny side and 
presenting densities below 100 kg~m$^{-3}$ (Kelley {\it et al.}, 2013). The velocity
of the individual icy grains is smaller than the expansion 
velocities expected for 1 $\mu$m pure water ice particles, confirming that
they are probably components of aggregates (Protopapa {\it et al.}, 2014).
\nocite{ahepoxi2011,kelleyetal13,protopapa2014}

Similarities between the material excavated from Tempel~1 and the outgassing
from Hartley~2 suggest icy aggregates built 
of relatively pure icy grains with sizes of about one micron. Quite comparable
values are also derived for near-infrared remote 
observations of the coma of comet 17P/Holmes after its October 2007 outburst,
with solid particles in the coma suggested to 
consist of both refractory dust and cold ice grains, not in thermal contact
(Yang {\it et al.}, 2009; Beer {\it et al.}, 2009).
\nocite{yajebu09}

\subsection{Results from Remote Observations}
\begin{figure}
\centering
\includegraphics[scale=0.3]{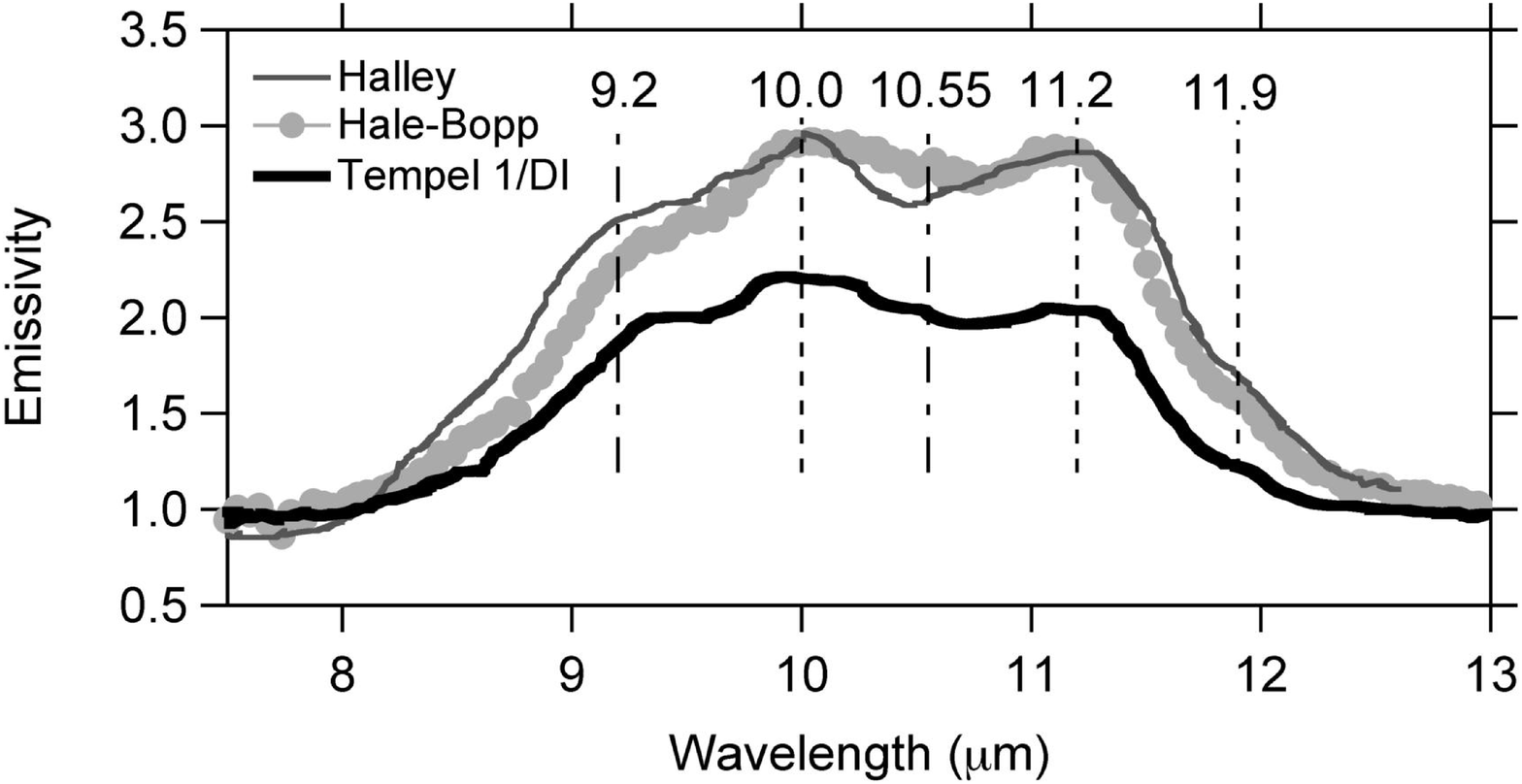}
\caption{Comparison of silicate features, as observed for 1P/Halley and
C/1995 O1 Hale-Bopp (adapted from Hanner, 1999), and for 9P/ Tempel 1 after
the Deep Impact event and release of subsurface material (adapted from Kelley
and Wooden, 2009). In each case the total flux is divided by the black body
contribution. The main features are at 10 $\mu$m, 11.2 $\mu$m and 11.9 $\mu$m
for crystalline olivine, and at 9.2 $\mu$m and 10.55 $\mu$m for crystalline
ortho-pyroxene.}\label{dust_spectrum}
\end{figure}

Solid particles in cometary comae have been studied for several tens of comets
bright enough to be observed from Earth or near-Earth-based observatories.
Results stem from infrared spectroscopic
observations (in the 8--13 $\mu$m atmospheric window or 
over a larger range from space observatories), as well as from linear
polarimetric observations (including polarimeric imaging 
techniques). After the results from Halley flybys and before space missions to
JFCs, the very active Oort cloud comet C/1995~O1 
(Hale-Bopp) increased our understanding of the composition of the solid component
of comae. Its activity was higher than for all other 
comets previously observed, and significant similarities in the IR dust spectrum
were found with comet Halley (Figure~\ref{dust_spectrum}; Hanner, 1999).
\nocite{hanner99}

\subsubsection{Infrared Spectroscopy: Silicates} 
Silicate components of the dust contribute to a weak continuum emission and
strong resonances in the 10 $\mu$m domain for Hale-Bopp. 
Different silicates were identified in spectra: amorphous pyroxene, amorphous
olivine and crystalline olivine. Closer to the Sun, 
Mg-rich crystalline pyroxene was also identified. The very significant strength
observed for the features in C/1995~O1 (Hale-Bopp) 
could be due to a large number of submicron-sized grains (0.2 $\mu$m or
smaller). The smaller Mg-rich crystals could come from 
the jets, although no differences in silicate mineralogy were observed between
the different regions, e.g., inner coma, coma and 
jets (Hayward {\it et al.}, 2000). Other Oort cloud comets present similar spectra
with Mg-rich grains (Hanner, 1999; Wooden {\it et al.}, 2004), but with lower 
strength of the features.
\nocite{hanner99,hahase2000,wowoha04}

Comet 9P/Tempel~1 was extensively observed before and after the Deep Impact
event.  Before impact, spectra contained a broad 
feature in the 8-12 $\mu$m range, attributed to amorphous silicates. Spectra
of JFCs with low dust production rates
are generally found to have weak silicate features, discernible in high
signal-to-noise Spitzer IRS spectra of comets (Kelley and Wooden, 2009).
These weak silicate features, attributable to the coma dust population, are
better revealed after subtraction of the relatively strong thermal contribution
from the nucleus (Hanner 1999; Sitko {\it et al.}, 2004; Harker {\it et al.},
2005; Harker {\it et al.}, 2007; Kelley and Wooden, 2009; Woodward {\it et al.},
2011).  Weak silicate features arise from a deficiency of small grains rather
than a lack of silicate material (Hanner 1999; Woodward {\it et al.}, 2011)
{\it et al.}, 2005; Kelley and Wooden).
After impact on Tempel~1, spectra showed a very
significant feature due to crystalline Mg-silicate
(see Figure~\ref{dust_spectrum}), 
similar to those observed for Hale-Bopp (Harker {\it et al.}, 2005; 2007; 
Lisse {\it et al.}, 2007). These features progressively disappeared as 
the plume dissipated away from the nucleus. Grains first ejected from Tempel~1
were mainly submicron-sized amorphous-carbon 
grains and submicron-sized silicates (Sugita {\it et al.}, 2005; Harker {\it et al.}, 2007).
The second part of the ejecta mainly contained
Mg-rich crystal silicates with submicron sizes, originating in the sub-surface
layer. It may be added that hydrated minerals, which have 
implications for the location of their formation region (Lisse {\it et al.}, 2007;
Kelley and Wooden, 2009) have also been detected. 
Amorphous carbon (submicron-sized) grains were suggested to come from a
carbonaceous layer formed by cosmic irradiation 
during the extremely long time spent by the nucleus in the transneptunian
region (Sugita {\it et al.}, 2007; Furusho {\it et al.}, 2007). This hypothesis is 
compatible with the observations of the impact crater by the Stardust-NExT 
mission (Schultz {\it et al.}, 2013). 
\nocite{hanner99,haetal05,kewo09,haetal05,haetal07,lisseetal2007}
\nocite{SugitaLPI,scheve2013,furushoetal2007,sitkoetal2004,woodwardetal2011}

Both Tempel~1 and Hale-Bopp have high-crystalline silicate fractions, but there
are important differences in their precise 
mineralogy (Lisse {\it et al.}, 2007; Kelley and Wooden, 2009). In comet Hale-Bopp
and many Oort cloud comets, only Mg-rich grains 
are present. Mg-Fe crystals are also found in Tempel 1 and JFCs. The presence
of Mg-silicates requires significant processing in 
the protosolar nebula (Wooden {\it et al.}, 2002). Annealing by heating may be an
important process to convert amorphous grains to 
crystals. New observations and modeling are necessary to understand the
transport and formation of the different classes of comets.
\nocite{lisseetal2007,kewo09,wooden2002}

\subsubsection{Infrared Spectroscopy: Organics and Icy Particles}
Refractory organics, which mainly contribute to the featureless emission and
possibly to a 3.4 $\mu$m emission (Green {\it et al.}, 
1992), are more difficult to detect. They can be mixed with rocky components,
and constitute a matrix gluing the different minerals, 
as in IDPs (Flynn {\it et al.}, 2003, 2013; Flynn, 2011). 
\nocite{flyetal2003,flyetal2013,fl2011}
They can also coat the dust grains, a process possibly having
taken place in the interstellar medium (Hanner and 
Bradley, 2004; Greenberg and Hage, 1990; Ehrenfreund {\it et al.}, 2004). When
particles are ejected in the coma, they get heated, 
leading to changes in their optical properties and even to evaporation and
destruction. Distributed sources may appear (Bockel{\'e}ee-Morvan {\it et al.},
2004; Cottin {\it et al.}, 2004) and fragmentation of dust particles
is likely to occur.
\nocite{greenetal92,habr04,grha90,ehetal03,bocrmuwe2004,cogabera04}

Remote detection of water ice is extremely difficult, with an icy grain halo
likely to be limited to a few hundred kilometers at 
heliocentric distances below 2.5 au (Hanner, 1981, Beer {\it et al.}, 2006, 2008),
\nocite{beetal06,beetal08}
meaning that high spatial
resolution is required to attempt any detection. 
Water ice has three absorption bands in the near-infrared, centered at 1.5,
2.0, and 3.0 $\mu$m. First detections of water ice were 
obtained on comet Hale-Bopp from the UKIRT telescope with the comet at about
7 au from the Sun (Davies {\it et al.}, 1997) and from 
the ISO space observatory at 2.9 au (Lellouch {\it et al.} 1998). Infrared spectra
of comets Hale-Bopp and C/2002~T7 (LINEAR) were 
reproduced with an intimate mixture of water ice and silicate spherical grains,
assuming water ice to be in an amorphous state 
(Davies {\it et al.}, 1997; Kawakita {\it et al.}, 2004).
In the infrared spectra of comets 17P/Holmes in its outburst and C/2011~L4
(PanSTARRS), water ice bands were detected at 2.0\,$\mu$m and reproduced with
submicron icy grains (Yang et al., 2009; 2014).
Remote observations of Tempel~1 from the
XMM-Newton observatory provided evidence 
for ice particles in the inner coma after the impact
(Schulz {\it et al.}, 2006).
Observations of color changes after the Deep Impact 
encounter in the visible and near IR domains also suggested an increase in
small grains and in ice relative to refractory dust in the 
coma (Knight {\it et al.}, 2007; Schleicher {\it et al.}, 2006). Indeed, remote observations from the Spitzer
observatory suggested a very high ice-to-dust ratio of 
about 10 in the excavated material (which greatly exceeds the gas-to-dust
production rate ratio of about 0.5 measured for the 
background coma), although a ratio in the 1 to 3 range cannot be excluded if a
large amount of material fell back to the surface 
and sublimated (Gicquel {\it et al.}, 2012).
\nocite{ha81,daetal97,lellouchetal98,kawakitaetal2004,schulz2006}
\nocite{knightetal2007,gicqueletal12,yajebu09,yakemeowwa14,scbaba06}

\subsubsection{Color and Polarimetric (Linear and Circular) Observations}
\label{polarization}
Observations of solar light scattered by solid particles in cometary comae
depend on the observational conditions (e.g., phase angle, wavelength) and
on the particles' physical properties (e.g., complex refractive indices and
composition, size, shape, geometric albedo). Differences in physical
properties between the particles in different regions of the coma are pointed
out by differences in the linear polarization of the scattered light and by
spectral variations in brightness and polarization. At visible wavelengths,
grains or particles with sizes of between
0.5 and 10\,$\mu$m seem to dominate the scattered light, affecting the degree
of linear
polarization and color variation in the brightness and in the polarization.

The color of the dust is associated with the variation of the scattered
intensity with
wavelength. In the visible domain, the scattered light is generally 
redder than the solar continuum (Jewitt and Meech, 1986; Kolokolova {\it et al.},
2004). The reddening slope decreases towards the 
near-infrared.
The color depends on the size distribution of the grains and
aggregates and on their refractive indices, mainly for 
grains larger than the wavelength. 
Some local 
variations of colors have been observed in comae. In comet Hale-Bopp, the color
in the curved jets was less red than the average 
background, possibly because of an increase in the number of submicron-sized
grains (Furusho {\it et al.}, 1999). At the beginning of 
the disruption of comet C/1999~S4 (LINEAR), the color was bluer while fast
moving small grains were detected, although the color 
of the largest fragments was red and became less red when the fragmenting
particles went farther away from the nucleus (e.g., 
Hadamcik and Levasseur-Regourd, 2003a). To interpret the slightly bluer color
in the visible and near infrared in the dust ejecta 
after the Deep Impact event, small submicron-sized grains were suggested, as
well as the presence of pure or mixed water ice 
crystals or the sublimation of organics (Hodapp {\it et al.}, 2007;
Fernandez {\it et al.}, 2007; Beer {\it et al.}, 2009). 
\nocite{beetal09,hale03a,hale09,hadamciketal14,kiselevetal2004}
\nocite{jeme86,kohalegu04,furushoetal99,hale03a,hodappetal07,fernandezetal07}
Generally, bluer color requires relatively transparent grain materials
such as water ice, Mg-rich silicates, or perhaps non-radiation damaged nor
heated organics (Kiselev {\it et al.}, 2004: Hadamcik and Levasseur-Regourd,
2009; Zubko {\it et al.}, 2011, 2012; Hadamcik {\it et al.}, 2014). 
\nocite{zubkoetal2011,zubkoetal2012}

The solar light scattered by cometary comae is partially linearly polarized.
Polarimetric properties vary with size and size distribution, morphology and
structure, as well as with complex refractive indices, of the scattering
particles. Observations can thus point out changes in dust properties
and provide information on composition through numerical and experimental
simulations (see e.g. for reviews, Kolokolova {\it et al.}, 2015).
Evidence for changes in local polarization was 
established by OPE/Giotto during the Halley flyby, with an increase in jets 
and a decrease near the nucleus at constant large phase angle 
and wavelength (Levasseur-Regourd {\it et al.} 1999).
Such behavior could also be monitored by
remote CCD imaging polarimery on comet Levy 
(Renard {\it et al.}, 1992) and later on other comets at different phase angles
(e.g., Jones and Gehrz, 2000; Hadamcik and Levasseur-Regourd, 2003b,
Hadamcik {\it et al.}, 2013; Deb Roy {\it et al.}, 2015;
Hines {\it et al.}, 2014).
\nocite{debroyetal15,hinesetal14,joge2000,kohole2015}

\begin{figure}
\centering
\includegraphics[scale=0.5]{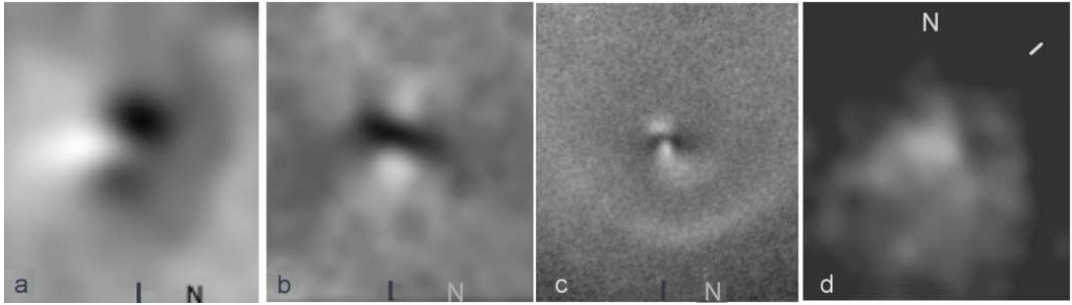}
\caption{Evidence for different properties of solid particles in different
regions of cometary comae, from linear polarization (P) maps at a given
phase angle ($\alpha$)
a. Comet C/1990 K1 Levy at $\alpha$ = 18$^\circ$, in a
field of view of 4000 km by 4000 km (black, P=-2.6\%; white, P= 0.5\%).
b. Comet C/1995 O1 Hale-Bopp at $\alpha$ = 7$^\circ$, in a field of 
38000 km by 38000 km (black, P=-6\%; white, P=2\%).
c. Comet Hale-Bopp, now at $\alpha$ = 44$^\circ$, in a field of
82000 km by 82000 km (black, P= 9\%; white, P= 18\%).
d. Comet 9P/Tempel 1, before Deep Impact event, without any conspicuous
feature, at $\alpha$ = 41$^\circ$, in a field of 70000 km x 70000 km
(black, P= 1\%; white, P=10\%). The solar direction is indicated by a black
or white tick.}\label{dust_morph}
\end{figure}

Three 
main regions of polarization have been detected, as illustrated in
Figure~\ref{dust_morph}:
the background coma, jet-like features (in
active comets) with a higher polarization than in the 
coma, and (for some comets only) a circumnucleus polarimetric halo, clearly
detected on polarimetric images of comets Levy and 
Hale-Bopp at different phase angles.
The polarization in the halo, if any, is more negative
than the surroundings for phase angles smaller
than about 20$^\circ$ (see Fig.~\ref{dust_morph}, a and b) 
and is lower than the surrounding coma for larger phase angles
(see Fig.~\ref{dust_morph}, c and d).

\begin{figure}
\centering
\includegraphics[scale=0.5]{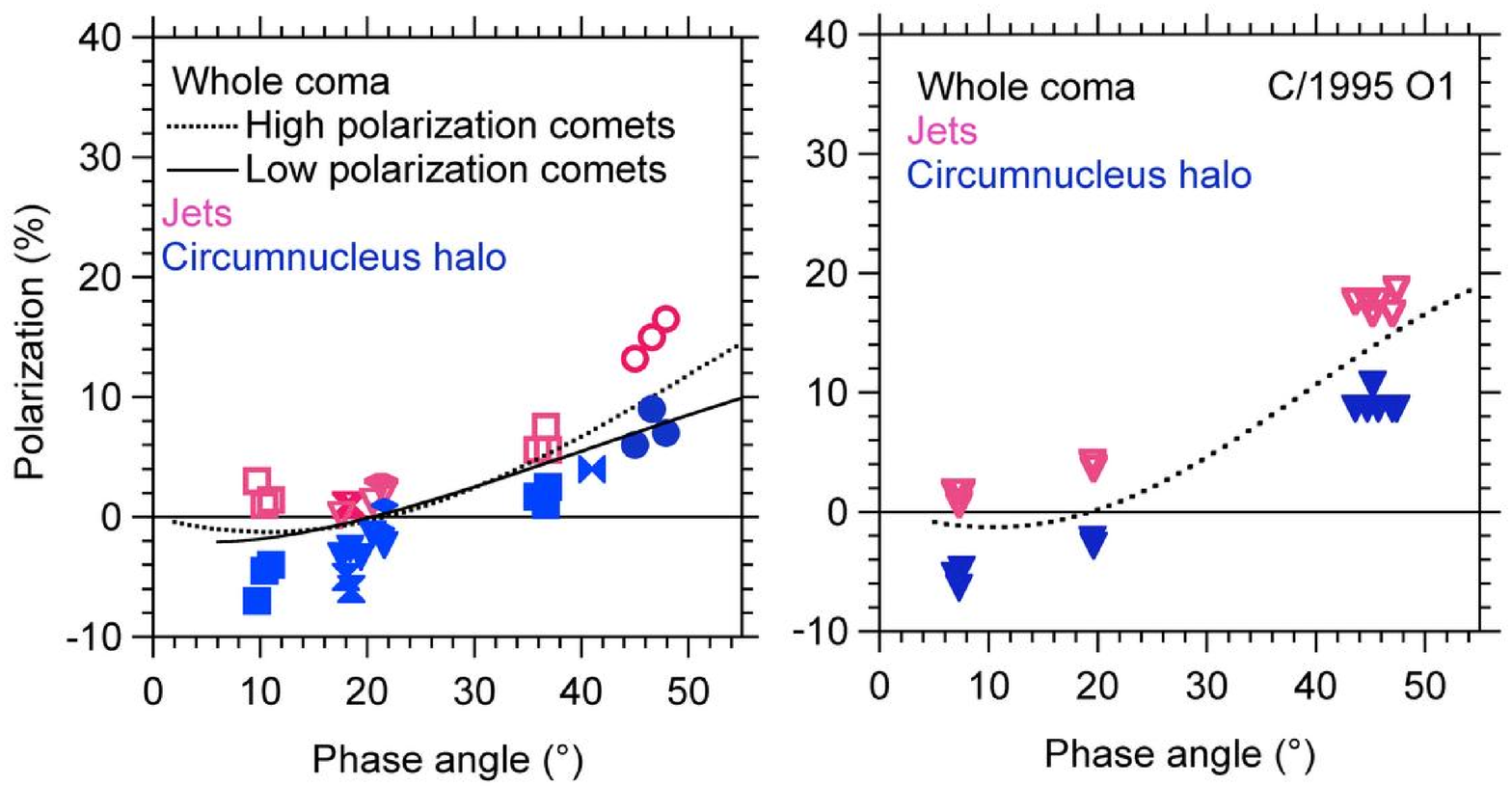}
\caption{Evidence for changes in linear polarization, and thus in solid 
particle properties, within specific coma features (jets, circumnucleus
polarimetric halo) observed at different phase angles.
Left, comets with
either high or low maxima in polarization; right, comet C/1995 O1
Hale-Bopp. While synthetic phase curves can be derived for whole cometary
comae, the polarization is higher in jets (unfilled symbols) and lower in the
circumnucleus halo (filled symbols), pointing out changes in composition
and sizes (Updated from Hadamcik and Levasseur-Regourd, 2003b).}
\label{dust_polar}
\end{figure}

The averaged polarization, 
measured through different increasing aperture sizes (e.g., Kiselev {\it et al.},
2001; Hadamcik {\it et al.}, 2013), provides the whole coma 
polarization when using large apertures (Hadamcik and Levasseur-Regourd, 2003b),
for a given phase angle and wavelength. 
Polarimetric phase curves show a small negative branch (down to about -1 to
-2\%), and inversion angle near 20$^\circ$, and
a broad positive branch with a maximum (up to about 15 to 
30\%) by 90$^\circ$ to 100$^\circ$ (e.g. Hadamcik and Levasseur-Regourd, 2003b).
From synthetic phase curves
built for all observed comets in the same wavelength 
range, two classes may be characterized (see Figure~\ref{dust_polar}):
the active comets with generally well
focused jets that present a high maximum in 
polarization, and the comets that present a lower maximum in polarization
(Levasseur-Regourd {\it et al.}, 1996). The latter are less 
active or may show an important decrease of polarization away from the
photometric center, such as comet Tempel~1 before the 
Deep Impact event (Hadamcik {\it et al.}, 2007a) or comet C/1996~Q1 (Tabur)
(Kiselev {\it et al.}, 2001).

For a given phase angle above 20$^\circ$, the whole coma polarization slightly
increases with the wavelength, at least up to about 2\,$\mu$m.
It seems to decrease for
wavelengths greater than 2\,$\mu$m, possibly due to 
some thermal emission (Kolokolova {\it et al.}, 2004), although
observations and thermal models of cometary comae show that the thermal
emission has its onset between 3\,$\mu$m and 4\,$\mu$m for
heliocentric distances in the 1--3\,{\sc au} range (Sitko {\it et al.}, 2004;
Harker {\it et al.}, 2002, 2004). A decrease of polarization with
increasing wavelength in the visible
has been observed for some comets such as 21P/Giacobini-Zinner 
(Kiselev {\it et al.}, 2000) or during some event such as the
complete disruption of comet C/1999~S4 (LINEAR) (Hadamcik {\it et al.}, 2003a).
In the circumnucleus polarimetric halo region, such a spectral gradient
inversion
was noticed through {\it in situ} observations of Halley 
and remote observations of Hale-Bopp, possibly providing clues to different
properties for freshly ejected dust particles.
\nocite{lemchafu99,reledo92,hale03b,kihoroko01,hadamciketal13,lehare96}
\nocite{kihoroko01,kohalegu04,kiselevetal2000}
\nocite{hawowoli02,hawowoli04}

In addition to its linear polarization, the light scattered by cometary dust
appears to present a weak circular polarization signal. The 
origin of circular polarization is still debated but could arise from the
alignment of dust particles, light scattering by asymmetric 
particles, or the presence of prebiotic homochiral organic molecules in
cometary dust. More observations are necessary to better 
constrain its origin. A discussion can be found in Kiselev {\it et al.}, 2014.
\nocite{kiroleko2014}

\subsubsection{Properties of Solid Particles Inferred from Light Scattering Measurements, with Emphasis on Composition}
Numerous authors have tried to constrain the scattering properties of cometary
dust. From experimental simulations, Hadamcik {\it et al.} (2007b) established
that
fluffy Mg-Fe-SiO and C mixtures could represent satisfactory cometary dust
analogs. From numerical simulations of polarization data and silicate emission
features, Kolokolova {\it et al.} (2007) concluded that the dust in comets with a
high maximum in polarization consists of highly porous aggregates
that may be associated with fresh
dust as in new comets.
The dust in comets with a lower maximum polarization
consists of less porous particles which may be associated with more
highly processed dust such as expected for the surfaces of JFC comets.
Compact porous particles also dominated the coma of long period comet
C/2007~N3 (Lulin), as evidenced by a negative branch that extended to
near-IR wavelengths (Woodword {\it et al.}, 2011).
Finally, whenever polarimetric phase curves are obtained on a large enough
range of phase angles in different wavelengths (as possible mostly for Halley
and Hale-Bopp), numerical and experimental
simulations may be used to infer some average
properties, such as size and size distribution, ratio between transparent and
absorbing materials, and ratio between fluffy aggregates and compact particles
(Levasseur-Regourd {\it et al.}, 2008; Lasue {\it et al.}, 2009;
Zubko {\it et al.}, 2012; Hines {\it et al.}, 2014). 
\nocite{hadamciketal07b,kokikiro2007,lalehaal09,hinesetal14,lezola2008}

The circumnucleus polarimetric halo provides an example of the complexity of
interpretations. First of all, it is not always observed; it can be hidden by
jets or not continuously present or may not exist at all. Secondly, its
interpretation is controversial. The deep negative branches observed for
C/1995~O1 (Hale-Bopp) and 81P/Wild~2 seem to be possible only if
transparent particles
exist in this region. Suggestions are large agglomerates of water ice grains,
an excess of Mg-rich silicate grains as compared to dark carbonaceous
compounds, and
relatively transparent organics, possibly covering silicate grains 
(Flynn {\it et al.}, 2003; Flynn {\it et al.}, 2013).
\nocite{flyetal2003,flyetal2013}
These
organics, heated in the coma after ejection, may become darker by
carbonization and
can release gases as an extended source (Hadamcik {\it et al.}, 2014).
Various compositions and evolutions
may actually co-exist in the innermost comae.
Additionally, a smaller polarization
may originate in some depolarization induced by
multiple scattering, as was certainly the case in the
plume ejected from Tempel~1 after impact. Nevertheless, the polarimetric
spectral gradient, possibly negative or neutral near the nucleus, and positive
after an evolution of the composition, could provide further clues.

There can be a strong synergy between the dust particle properties derived
from studies of color and linear polarization with their properties assessed
by studies of coma dynamics and thermal emission, where comet Hale-Bopp is an
excellent example.  As stated in subsection~\ref{polarization}, 
Hale-Bopp’s curved jets or arcs were less red than the average background,
possibly because of an increase in the number of submicron-sized grains. 
Compared to the background coma, Hale-Bopp arcs also showed higher linear
polarization (Jones and Gehrz, 2000), higher dust color temperatures
(Hayward et al., 2000), strong silicate features (Hayward et al., 2000)
indicative of a high concentration of submicron silicate crystals as well as
micron-sized porous particles (Harker et al., 2002, 2004). The coma
dynamical models showed that arc/jet particles needed to be small (submicron)
and relatively transparent in order that solar radiation pressure not smear
out the arcs nor compress the arc spacing (Hayward et al., 2000).
\nocite{joge2000}
Ground
truth, as expected from a rendezvous mission such as Rosetta, monitoring the
dust properties of a given comet at different distances from the nucleus and
from the Sun, coupled with remote observations, is certainly  of major
importance for future interpretations of spectroscopic and polarimetric
observations of various comets.

\subsection{Sun-Grazing Comets}
When a comet gets very close to the Sun, it is heated sufficiently
that the refractory particles can start to sublimate,  releasing
metals such as Na and K in the coma. Observations of these metals
can give key insights into the atomic abundances in cometary dust 
that can only otherwise be gleaned with dust analyzers {\it in situ}.
However, these lines only become prominent when a comet is
very close to the Sun, challenging our ability to observe the comet
from the ground or with space telescopes owing to the small solar
elongation. 
Indeed, these comets are so close to the Sun that the observations must
be obtained during the daytime.
As noted above, comet C/2012~S1 (ISON) was one such
Sun-grazing comet that was extensively observed, though it
faded substantially when closest to the Sun, limiting our ability
to detect the metal lines.

Comet C/1965~S1 (Ikeya-Seki) was one of the brightest
Sun-grazing comets to be studied spectroscopically.  
Observations were obtained
from Kitt Peak National Observatory (Slaughter, 1969;
Arpigny, 1979),
Lick Observatory (Preston, 1967), Sacramento Peak (Curtis, 1966),
Haute Provence (Dufay {\it et al.}, 1965), Radcliffe (Thackeray {\it et al.}, 
1966), and possibly others, 
when the comet was as close to the Sun as $\sim14$ solar radii.
\nocite{arpigny1979,slaughter1969,preston1967,curtis1966,dufay1965,
thackeray1966}

In addition to the normal cometary gas emissions, emission lines due
to Na~I, K~I, Ca~I, Ca~II, Cr~I, Mn~I, Fe~I, Ni~I, Cu~I, Co~I, and V~I
were observed.
From analysis of the Fe lines, a Boltzmann temperature of $\sim4500\,K$
was determined (Slaughter, 1969; Preston, 1967).
Arpigny (1979) noted that Mg, Al, Si ad Ti were not detected.
Compared with solar abundances, the abundances of some of these metals were 
underabundant, while others were overabundant (Arpigny, 1979; Preston, 1967).
However, arguments were presented (Arpigny, 1979) that the 
relative {\it elemental} abundances were probably close to solar or
meteoritic values.

\subsection{Dust Composition: Rosetta Capabilities and Early Results}
The instruments on the Rosetta mission that could provide direct information
on the composition of solid particles for 67P/Churyumov-Gerasimenko (CG)
are mostly COSIMA and VIRTIS 
on the main Rosetta spacecraft, to be used during the whole rendezvous mission,
and COSAC at Philae landing. The COmetary 
Secondary Ion Mass Analyser, COSIMA, is the first instrument applying SIMS
(secondary ion mass spectrometry) technique to {\it in situ} analysis.
It collects dust particles at the Rosetta location on targets
handled by a manipulation unit, in order to obtain time-of-flight spectra over
a mass range from 1 to 3500 amu, leading to the composition of the dust
particles, whether they are organic or 
inorganic (Kissel {\it et al.}, 2007). The Visual IR Thermal Imaging Spectrometer,
VIRTIS, provides imaging spectroscopy in the 
0.25-5.0 $\mu$m spectral range, in order to map composition and evolution of
dust jets in the coma, as well as to derive the 
composition of the dust grains in the inner coma and on the surface (Coradini
{\it et al.}, 2007).  By August 2014, strong hints of carbon-bearing 
compounds, with spectral features compatible with complex macromolecular
carbonaceous materials, were already noted on the nucleus surface.
(http://blogs.esa.int/rosetta/2014/09/08/virtis-maps-comet-hot-spots/).
The COmetary 
SAmpling and Composition experiment, COSAC, relies on a multi-column
enantio-selective gas chromatograph, coupled to a linear 
time-of-flight mass spectrometer to derive the composition of volatile species
collected at the surface and at about 20 cm below the surface
with the Philae lander (Goesmann {\it et al.}, 2007).
\nocite{kietal2007,2007SSRvCoradini,2007SSRvGoesmann}

Interpretation of data, e.g., from dust experiments GIADA (Colangeli
{\it et al.}, 2007), COSIMA (Kissel {\it et al.}, 2007)
and MIDAS (Riedler {\it et al.}, 2007) on Rosetta, as 
well as from CONSERT (Kofmann {\it et al.}, 2007), APSX (Klingelh{\"o}fer
{\it et al.},
2007) or Ptolemy (Wright {\it et al.} 2007) on Philae, should 
also significantly contribute to a better understanding of the dust
composition, and on its evolution i) from the subsurface to the 
surface of the nucleus and to the coma and ii) from large and decreasing solar
distances before perihelion passage to increasing solar 
distances after perihelion passage.
\nocite{2007SSRvColangali,2007SSRvRiedler,2007SSRvKofman,2007SSRvKlingelhofer,2007SSRvWright}

Preliminary results from the Rosetta mission have already established that
the surface of CG is rich in non-volatile organics, likely complex mixtures
of various carbon-hydrogen (aromatic and aliphatic), oxygen-hydrogen
(carboxylic or alcoholic) and nitrogen-hydrogen groups (Capaccioni {\it et al.},
2015). The largest particles (with sizes greater than 50\,$\mu$m) are
demonstrated to be fluffy and fragile aggregates, and found to present
sodium-rich surfaces (Schulz {\it et al.}, 2015). 
While the presence of such aggregates had been suspected from simulations 
(e.g., Hadamcik et al., 2007a; Levasseur-Regourd et al., 2008), 
the Rosetta ground-truth now allows us to speculate about the early Solar 
System evolution. A significant amount of non-volatile organics could have 
enriched the surface of terrestrial planets, with fluffy particles more 
resistant to atmospheric ablation than compact ones.
More results and
discoveries are soon expected, from Rosetta pre-perihelion (at solar distances
less than 3\,{\sc au}) and post-perihelion measurements, and from
Philae measurements.
\nocite{cacofi15,schulzetal2015,hadamciketal07a}

\section{Summary}
We have learned a great deal about comets from the studies outlined in
this paper.  However, the picture is not yet fully coherent since 
we find different stories when we change wavelength.  The Rosetta
mission will help us to gain detail and to compare with the data
that have come before.  However, it is obvious that all that can
be learned from the enigmatic comets has not yet been found.

\noindent
\underline{Acknowledgements:} The authors would like to thank Dr. Paul Feldman,
Dr. Jacques Crovisier, Dr. Neil Dello Russo and Dr. Adam McKay for helpful
comments and making data available for our use.
Partial support from CNES (the French spatial agency) is acknowledged.
\newpage

\end{document}